# Validation of contact mechanics models for Atomic Force Microscopy via Finite Elements Analysis and nanoindentation experiments

L. Dal Fabbro, H. Holuigue[1], M. Chighizola[2,]* and A. Podestà*

Dipartimento di Fisica "Aldo Pontremoli", Università degli Studi di Milano, via G. Celoria 16, 20133, Milano, Italy.

CIMaINa, Università degli Studi di Milano, via G. Celoria 16, 20133, Milano, Italy.

*Corresponding authors. E-mail: matteo.chighizola@ifom.eu, alessandro.podesta@unimi.it

## ABSTRACT

In this work, we have validated the application of Hertzian contact mechanics models and corrections for the analysis of force vs indentation curves, acquired using spherical indenters on linearly elastic samples, by means of finite elements simulations and AFM nanomechanical measurements of polyacrylamide gels possessing a thickness gradient. We have systematically investigated the impact of both large indentations and vertical spatial confinement (bottom effect) on the accuracy of the nanomechanical analysis performed with the Hertz model for the parabolic indenter compared to the Sneddon model for the spherical indenter. We demonstrated the accuracy of the combined correction of large indentation and bottom effects for the Hertz model proposed in the literature in the framework of linearized force vs indentation curves acquired using spherical indenters, as well as a validation of a new linearized form of the Sneddon model. Our results show that the corrected Hertz model allows to accurately quantify the Young's modulus of elasticity of linearly elastic samples with variable thickness at arbitrarily large indentations.

[1] Present affiliation: Mechanobiology Institute (MBI), National University of Singapore, 5A Engineering Drive 1, 117411, Singapore

[2] Present affiliation: Institute FIRC of Molecular Oncology (IFOM), Via Adamello 16, 20139 Milan, Italy.

# 1. INTRODUCTION

The micro- and nanoscale characterisation of the mechanical properties of systems and device components is of increasing importance in several fields like biology, where the elastic properties of cells, tissues, and extracellular matrix (ECM) can affect the behaviour and fate of an organ[1–6], or medical engineering, where microdevices with specific elastic properties are employed[7,8].

Atomic Force Microscopy (AFM) is an ideal tool for the quantitative and non-destructive characterisation of the mechanical properties of biological and non-biological samples at the sub-micrometre scale[9–11]. AFM provides high spatial and force resolution and is very versatile, including the ability to work in physiological solution and controlled environments, the freedom of choosing the best tip dimensions and geometry to match the typical length scales of the system under investigation[12–14].

While sharp AFM tips are mandatory when high spatial resolution is necessary, spherical probes (colloidal probes, CPs) possess several characteristics that make them suitable for the investigation of mechanical properties of soft or biological samples[14]. Indeed, a well-defined interaction geometry (sphere on flat, sphere on sphere) and reduced stress and strain in mechanical tests make the application of contact mechanics models more reliable, and the interpretation of results less ambiguous; the selection of the tip radius to match the characteristic length scales of the system under investigation provides better averaging of the mechanical signals together with a system-adapted spatial resolution[13,15].

A simple contact mechanics model describing the deformation of elastic solids was proposed by Heinrich Hertz in 1882[16,17]. The Hertz model can be used to describe the indentation of a purely elastic, semi-infinite half space under the pressure exerted by a paraboloidal indenter. It is well known that the application of the Hertz model to real systems is based on a series of assumptions and, in some cases, gross approximations: the sample must be uniform and isotropic; interfacial adhesion must be absent; the strain and stress must be sufficiently small to ensure a linear elastic response; the sample must not exhibit a constrained mechanical response due to its spatial confinement and/or finite dimensions (the Bottom Effect problem[18–21]). In addition, to apply the Hertz model to data collected using spherical indenters, the indentation $\delta$ of the probe must be small compared to the radius R of the tip ($\delta \ll R$), otherwise the parabolic approximation of the spherical profile will be inaccurate (the Large Indentation problem[22,23]).

Microscopic systems like single cells or thin tissue slices for histological analysis make the assumptions behind the use of the Hertz model to fit nanoindentation data hard to be respected. Cell height usually varies between 5-15 μm and its internal structure is extremely heterogeneous in all three dimensions, forcing to indent up to a few microns to characterise the overall mechanical response, and not just the elastic contribution of the cell membrane coupled to the actin cortex[15,24].

Achieving large indentions up to a few microns is even more critical for tissue or tissue-derived samples. Indeed, these samples are usually far more heterogeneous than cellular systems, across a broader range of length scales (from 50 nm to 100 μm), which requires probing a larger volume with the indenter. In addition, their surface may present micron-scale roughness due to the above-mentioned structural complexity and to the slicing process, which produces irregular interfaces, due to cell detachment and disentangled ECM fibres.

Excluding nonlinear effects, the consideration of which goes beyond the aims of the present work, the requirement of achieving large indentations on cells and other finite-thickness systems (including thin tissues or ECM slices) exposes to the risk of having both, the large indentation, and the bottom effect issues, significantly influencing the force vs indentation curves and greatly affecting the precision of the Hertzian analysis.

To mitigate the large indentation issue in the framework of the Hertz model, it is possible to increase the radius R of the tip to reduce the $\delta/R$ ratio; this measure however can limit the maximum indentation achievable or mandate the use of very rigid cantilevers, with consequent loss of force sensitivity, beside causing a severe loss of spatial resolution. Alternatively, the model developed by Sneddon to describe the indentation of an elastic half-space by a spherical indenter [25] can be used; this model (hereafter simply called the Sneddon model) does not suffer from the constraint $\delta \ll R$. Unfortunately, Sneddon's equation for the contact radius of the spherical probe cannot be cast in an analytic close form but requires numerical methods to be solved (see section 2.2).

A typical mitigation of the bottom effect typically consists of limiting the maximum indentation to a small fraction (well below 1/10) of the sample thickness, or height, which may keep from sensing the elastic contributions of the deeper layers of the system under investigation. To overcome the above-mentioned limitations of the Hertz theory, the Bottom Effect Correction (BEC) [18–20,26] and the Large Indentation Correction (LIC)[22,23] have been proposed. The aim of such corrections is to extend the applicability of the Hertz model for the paraboloidal indenter to those cases where the estimation of the Young's modulus (YM) is biased due to the bottom effect and the large indentations, respectively.

Despite the increasing occurrence of application of BEC in published reports (LIC is by far less considered), a systematic validation and a suitable integration of both correction methods in a single experimental and data analysis approach are still missing. This is also due to the difficulty in producing suitable test samples for the nanomechanical investigation allowing to probe the influence of the varying thickness, spanning the range from the strong spatial confinement to the bulk conditions. The scarce availability of reference samples for nanomechanical tests limits the standardisation of the experimental and data analysis protocols [27,28].

In this work, we have validated the application of Hertzian contact mechanics models and corrections in the framework of linear elasticity for the analysis of force vs indentation curves, acquired using spherical indenters, by means of finite elements simulations and AFM nanomechanical measurements of polyacrylamide gels possessing a thickness gradient. FEA is a suitable tool for the simulation of the mechanical response of materials subject to the stress applied by an indenter under controlled conditions[26,29–32]. We have used FEA to simulate force vs indentation curves on ideal elastic samples[29,30].

We here discuss the accuracy of the Hertz model[33,34] compared to the Sneddon model for the spherical indenter, both in bulk systems and in conditions of vertical spatial confinement and assess the accuracy of the existing large indentation and bottom effect corrections for the Hertz model. We present a simplified linearised version of the Sneddon model, which can be implemented efficiently in the same data analysis framework we developed for the use of the linearised Hertz model. Eventually, we demonstrate that LIC and BEC can be coupled in a single correction function for the Hertz model and provide an experimental validation of the proposed methods based on AFM nanoindentations measurements with colloidal probes on polyacrylamide test samples fabricated on purpose to possess a gradient of thickness.

## 2. THEORETICAL MODELS AND CORRECTIONS

We present here a concise summary of the theoretical models used in this work, under the assumption of linear elasticity. For a deeper presentation and discussion of contact mechanics models, the reader is referred to some recent review papers, see Refs [17,35]

## 2.1. Hertz model

The Hertz model describes the non-adhesive contact between two uniform, isotropic elastic bodies, within the framework of linear elasticity (small strain and stress), thus considering the deformations small compared to the dimensions of the bodies[36]. Originally developed for describing the elastic contact between two spherical bodies[16], the Hertz model strictly applies to the case of a paraboloidal indenter. It can be used to describe indentation by a spherical indenter provided the indentation is small compared to the tip radius ($\delta \ll R$); in this case the parabolic profile represents a fairly accurate approximation of the circular one. In the case of indentation by an infinitely rigid paraboloid of an elastic half space with Young's modulus E and Poisson's ratio $\nu$, the relation between the applied force $F_{\text{Hertz}}$ and the indentation $\delta$ is:

$$F_{\text{Hertz}} = \frac{4}{3}\frac{E}{1-\nu^2}\sqrt{R}\delta^{3/2} \tag{1}$$

where $R$ is the radius of curvature of the indenter, identified in apical region of the tip. The contact region has circular cross section, and a simple relation links the tip radius $R$, the indentation $\delta$ and the contact radius $a$ (Figure S1a, blue circles):

$$\delta = \frac{a^2}{R}. \tag{2}$$

From Eq. 2 it follows: $a = \sqrt{\delta R}$.

## 2.2. Sneddon model for the spherical indenter

Sneddon developed a solution[37] for the case of the non-adhesive indentation of an elastic half-space by a rigid spherical indenter, represented by the system of Eqs. 3 and 4:

$$F_{\text{Sneddon}} = \frac{E}{2(1-\nu^2)}\left[(a^2 + R^2)\ \ln\left(\frac{R+a}{R-a}\right) - 2aR\right] \tag{3}$$

$$\delta = \frac{1}{2}a\ln\left(\frac{R+a}{R-a}\right) \tag{4}$$

The Sneddon model for the spherical indenter (in the following text, simply called the Sneddon model) is not subject to the limitation $\delta \ll R$. The drawback of this model is the lack of an analytical solution linking the force to the indentation, since Eq. 4 cannot be inverted analytically to obtain the relation $a = a(\delta)$; hence, a numerical solution is required. It is worth noting that Eq. 2, valid for the Hertzian contact, represents the limit of Eq. 4 when $a/R \to 0$, which justifies the use of the simpler Hertz model when using a spherical indenter as long as the indentation $\delta$ is (very) small compared to the radius $R$ (see Figure S1a).

### 2.3. Large indentation corrections for the Hertz model

The Hertz model is simple (analytic and linearizable, see below), and it is therefore desirable to use it to fit indentation data obtained with spherical indenters and arbitrarily large indentations, typically up to $\delta = R$. To this purpose, in this work we have selected two different corrections, presented in the papers published by Kontomaris *et al.*[22] and by Muller et al[23]. Other corrections, not considered here, could be readily incorporated within the presented protocol. Both correction functions $1/\Omega_{K/M}$ depend on the nondimensional ratio $\gamma_R(\delta) = \delta/R$ and transform the experimentally measured force-indentation curve $F$ into an apparent Hertzian curve $F_{\text{Hertz}}$ (Eq. 1), which can be fitted by the Hertz model across the whole range of indentations:

$$F_{\text{Hertz}}(\delta) = \frac{F(\delta)}{\Omega_{K/M}(\gamma_R(\delta))} \tag{5}$$

The function $\Omega_K$ presented in Kontomaris *et al.*[22] is a power series in $\gamma$:

$$\Omega_K = c_1 + \sum_{k=2}^{6} \frac{3}{2k} c_k \gamma^{k-\frac{3}{2}} \tag{6}$$

with coefficients:

$$c_1 = 1.0100000$$
$$c_2 = -0.07303003$$
$$c_3 = -0.1357000$$
$$c_4 = 0.0359800$$
$$c_5 = -0.0040240$$
$$c_6 = 0.0001653$$

The function $\Omega_M$ presented in Muller et al[23] is a polynomial in $\gamma$:

$$\Omega_M = 1 - \frac{1}{10}\gamma - \frac{1}{840}\gamma^2 + \frac{11}{15120}\gamma^3 + \frac{1357}{6652800}\gamma^4 \tag{7}$$

Both correction functions act pointwise on the force curve, since their values depend on the indentation through the ratio $\gamma$ (see Figure S2a). The contact radii corrected for the large indentation effect are in excellent agreement with the Sneddon's radius (see Figure S1a and Note S1).

## 2.4. Bottom effect corrections for the Hertz model

The presence of a rigid substrate underneath the sample represents a boundary condition, which, in the case of a large indentation to thickness ratio, can cause strong perturbation of strain and stress fields and therefore influence the measured elastic modulus[17,28]. The bottom effect is a special case of the more general case of three-dimensional spatial confinement of the elastic body, where the strain and stress fields are constrained also laterally[38]. This could be relevant, for example, when cells are at confluence, tightly arranged within a nearly two-dimensional monolayer, and firmly connected through cadherin bonds.

We have selected two correction functions for the bottom effect, developed for the paraboloidal indenter, in the case of linear elastic response: the first one ($1/\Delta_D$) was developed by Chadwick, and published in Dimitriadis *et al.*[18]; the other correction function ($1/\Delta_G$) was developed by Garcia and published in Garcia *et al.*[19] (see Figure S2b). The correction presented in the paper by Long et al[26], which takes into account the vertical confinement for a non-linear neo-hookean material, as well as others, have not been investigated in this work, although they could be readily incorporated within the presented protocol. These corrections can be applied, similarly to the large indentation case, to transform the experimentally measured force-indentation curve $F(\delta)$ into an apparent Hertzian curve $F_{Hertz}(\delta)$, describing the case of an infinitely extended elastic half space across the whole range of indentations:

$$F_{\mathrm{Hertz}}(\delta) = \frac{F(\delta)}{\Delta_{D/G}(\chi_{R,h}(\delta))} \tag{8}$$

The functions presented by Dimitriadis *et al.*[18] for a paraboloidal indenter refers to the following boundary conditions (BCs): *i*) a sample bonded to the rigid substrate ($\Delta_D^{\mathrm{bonded}}$); *ii*) a sample that is allowed to slide over it ($\Delta_D^{\mathrm{unbonded}}$); *iii*) a sample that is partially bonded to the substrates ($\Delta_D^{\mathrm{mixed_BCs}}$, like for single cells[13]):

$$\Delta_D^{\mathrm{bonded}} = 1 + 1.133\chi + 1.283\chi^2 + 0.769\chi^3 - 0.0975\chi^4 \tag{9a}$$

$$\Delta_D^{\mathrm{unbonded}} = 1 + 0.884\chi + 0.781\chi^2 + 0.386\chi^3 + 0.0048\chi^4 \tag{10}$$

$$\Delta_D^{\mathrm{mixed_BCs}} = 1 + 1.009\chi + 1.032\chi^2 + 0.578\chi^3 - 0.0464\chi^4 \tag{11}$$

Following Garcia et al.[19], The sign of the 4$^{\mathrm{th}}$ order coefficient in Eq. 9a has been changed with respect to the original work of Dimitriadis et al. (the value of the 4$^{\mathrm{th}}$ order coefficient in Eq. 11 has been also changed, accordingly); the same corrections should be applied to Eqs. 2-4 of our previous work, Ref.[13]). In these equations, the driving nondimensional parameter is $\chi_{R,h}(\delta) =$ $\sqrt{R\delta}/h$, where $h$ is the height (or thickness) of the sample measured with respect to the hard substrate. The parameter $\chi$ represents the ratio between the Hertzian contact radius $a$ (see Eq. 2) and the sample height $h$, which suggests that bottom effects are negligible in the limit $a \ll h$ (like for sharp probes, even though a higher stress would be produced). Noticeably, the bottom effect depends on the ratio of horizontal ($a$) to vertical ($h$) dimensions, and not simply on $\delta/h$, which in turn reminds us that the elastic deformation of a body is a truly three-dimensional process. Moreover, this also suggests that large spherical tips are more prone to bottom effects than sharp pyramidal ones.

The function $\Delta_G$ presented by Garcia *et al.* [19] for a paraboloidal indenter on a sample bonded to the rigid substrate is:

$$\Delta_G = 1 + 1.133\chi + 1.497\chi^2 + 1.469\chi^3 + 0.755\chi^4 \quad (12a)$$

It is worth noting that the corrective factors of the BECs reported in Eqs. 9a-12a are computed based on the assumption of incompressibility of the sample, thus imposing a Poisson's ratio $\nu = 0.5$. Since the FEA simulations are performed for an elastic medium with Poisson's ratio $\nu = 0.49$, we report the corrective factors $\Delta_D$ and $\Delta_G$ for the bonded case with $\nu = 0.49$, truncated to the fourth order (Eq. 9b and Eq. 12b). While Dimitriadis *et al.*[18] report the equations to calculate the BEC coefficients for an arbitrary Poisson's ratio, in the work of Garcia *et al.* [19] explicit formulae are not present; from the equations reported by Garcia *et al.*, we derived approximate expressions for the BEC coefficients for an arbitrary Poisson's ratio (see Note S2 for details).

$$\Delta_{D,\nu=0.49} = 1 + 1.112\chi + 1.237\chi^2 + 0.712\chi^3 - 0.133\chi^4 \quad (9b)$$

$$\Delta_{G,\nu=0.49} = 1 + 1.112\chi + 1.444\chi^2 + 1.374\chi^3 + 0.645\chi^4 \quad (12b)$$

Based on Garcia's formulae, we also derived the expression of the contact radius in confined geometry for a spherical tip up to the 5$^{th}$ order in $\chi$, extending Garcia's 1$^{st}$ order result (see Note S2 and Figure S1b-d). The corrected contact radius is in good agreement with the predictions of finite elements simulations.

# 3. MATERIALS AND METHODS

## 3.1 Finite Elements Analysis

Finite elements simulations were performed using ANSYS Mechanical (ANSYS Student), the free version of ANSYS software, to produce ideal force vs indentation curves describing the mechanical response of elastic films in different regimes of spatial confinement, i.e., with different $\delta/R$ and $\sqrt{\delta R}/h$ ratios (Figure 1a). The large indentations option, which retains the quadratic terms in the deformation tensor[39], was used in the simulations, to secure convergence. The resulting material shows, within the mesh-related error, ideal elastic behaviour across the whole tested indentation range, for both bulk and thin film configurations, as witnessed by the linearity of the effective strain ($\varepsilon$*) vs effective stress ($\sigma$*) figures of merit (according to Tabor's definition[30]: $\varepsilon^* = 0.2a/R$, $\sigma^* = F/\pi a^2$ as well as to the similar Kalidindi's definition[40–42], see Figure S3, S4 and Note S3). The linearity of the simulated systems was further checked through the equivalent, local von Mises stress and strain (Figure S5 and Note S4).

We took advantage of the axial symmetry of the system, which allowed to create two-dimensional meshes and take full advantage of the reduced number of nodes available ($N_{max}$ = $1.28*10^5$). The simulated systems consist of a rigid spherical or paraboloidal tip with a wide range of tip radii $R$ indenting an elastic medium of different thickness $h$ to simulate both the bulk and the thin film regimes. The medium width $L$ was chosen to properly mimic the condition of an elastic medium infinitely extending in the lateral directions (see Table 1). The lateral dimensions of the simulated systems were chosen to be large enough to neglect spatial confinement-related effects[38]; the validity of this hypothesis was confirmed for each simulation by observing that the vertical and lateral displacements at the boundaries were negligible.

Four models were studied (see Table 1): Bulk-Sphere (B-S), Bulk-Paraboloid (B-P), Thin Film-Paraboloid (TF-P) and Thin Film-Sphere (TF-S); the first two models mimic the indentation of an infinite linearly elastic half space by a spherical and a paraboloidal indenter, respectively, whereas the latter two models mimic the indentation of a linearly elastic thin film by paraboloidal and spherical indenters, respectively. The material was simulated with nominal Young's modulus $E_{nom} = 0.5$ MPa and Poisson's ratio $\nu = 0.49$. The choice of $\nu = 0.49$ was made to simulate a nearly incompressible material[43] (the choice $\nu = 0.5$ keeps finite elements simulations from converging). In all models, the medium is bonded to a rigid substrate. In the range of applied forces, the tip deformation is negligible since its Young's modulus is six orders of magnitudes greater than that of the elastic material.

The ***B-S model*** (Bulk-Sphere) consists of a rigid spherical tip of radius $R = 5\ \mu m$ indenting a medium with thickness $h_{bulk} = 100R$. This model was used to investigate the large indentation effect. No bottom effect is expected.

The ***B-P model*** (Bulk-Paraboloid) consists of a rigid paraboloidal tip with radius of curvature $R = 5\ \mu m$ indenting a medium with thickness $h_{bulk} = 100R$. This model was used to investigate the large indentation effect. No bottom effect is expected.

The ***TF-P model*** (Thin Film-Paraboloid) consists of a rigid paraboloidal tip with radius of curvature $R = 5\ \mu m$ indenting a medium with thickness $h_{film} = 2R$. This model was used to investigate the bottom effect. No large indentation effect is expected for a parabolic indenter.

In the ***TF-S model*** (Thin Film-Sphere) indentation, thickness and tip radius are varied to investigate both large indentation and bottom effects. The details of the different thin film systems simulated are reported in Table 1.

Representative simulated FCs are shown in Figure 1b (512 points per single FC). Simulated FCs were fitted using the Hertz model with LIC and BEC and the Sneddon model for the spherical indenter. The maximum indentation in all the cases shown in Figure 1 was $\delta = 3\ \mu m$. It is possible to observe the good agreement with the theoretical models in all three configurations. Moreover, the impact of the indenter geometry and of the spatial confinement is negligible if the $\delta/R$ ratio is below 0.2. For larger $\delta/R$ values, irrespective to the indenter geometry, the bottom effect becomes dominant; it is evident that it is necessary to apply much higher forces to indent of the same amount a spatially confined material (compare TF-S with B-S and B-P). The Sneddon and the Hertz models represent the reference contact mechanics models to describe the indentation of an elastic half-space by spherical and parabolic indenters (B-S and B-P systems, respectively), irrespective to the value of the maximum indentation.

Comparing B-S and B-P systems, it can be noticed that when indentation increases, the force needed to obtain the same indentation with the parabolic indenter increases; this is due the difference in the contact area (see Figure S1a). Indeed, the parabolic contact area grows faster than the spherical one. The discrepancy between the simulated TF-S FC and the Hertzian FC corrected for the bottom effect is caused by the fact that the Hertz model is exact for the parabolic indenter, while the simulated curve assumed a spherical indenter. A suitable application of the large indentation correction is supposed to correct further this discrepancy (we will discuss this in Section 4.1.2).

| Model name | Tip shape | Tip Radius [$\mu m$] | Sample thickness [$\mu m$] | Sample width [$\mu m$] | YM [MPa] | Poisson ratio | Corrections to the Hertz model | Indentation [$\mu m$] |
|---|---|---|---|---|---|---|---|---|
| **B-S (Bulk-Sphere)** | Sphere | 5 | 500 | 120 | 0.5 | 0.49 | LIC | 5 |
| **B-P (Bulk-Paraboloid)** | Paraboloid | 5 | 500 | 120 | 0.5 | 0.49 | None | 5 |
| **TF-P (Thin Film-Paraboloid)** | Paraboloid | 5 | 10 | 120 | 0.5 | 0.49 | BEC | 2.5 |
| **TF-S (Thin Film-Sphere)** | Sphere | 5 | 10 | 120 | 0.5 | 0.49 | BEC, LIC, BEC+LIC | 3.1 |
| **TF-S (Thin Film-Sphere)** | Sphere | 10 | 8 - 60 | 200 | 0.5 | 0.49 | BEC, LIC, BEC+LIC | 1 |
| **TF-S (Thin Film-Sphere)** | Sphere | 1 - 65 | 20 | 50 - 400 | 0.5 | 0.49 | BEC, LIC, BEC+LIC | 1 |

Table 1. Summary of the different systems simulated for this study. In the last three rows, which correspond to the three Thin-Film systems, each of the three parameters, indentation $\delta$, height $h$ and tip radius $R$, were varied, while fixing the other two.

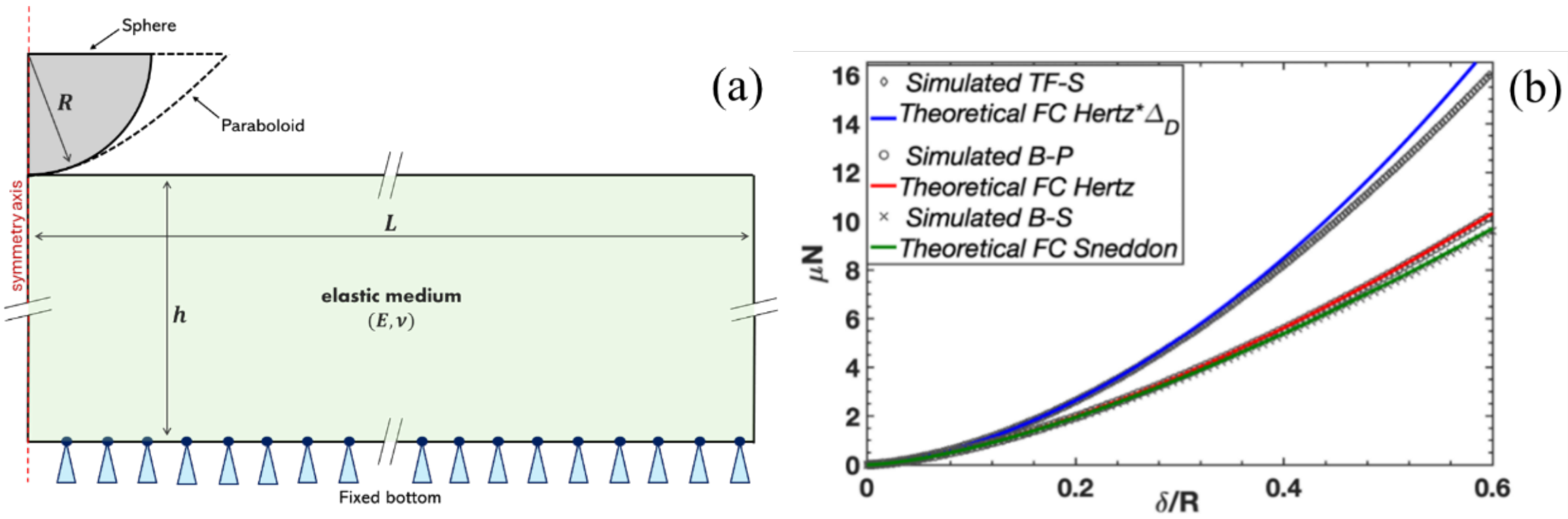

**Figure 1.** (**a**) Representation of the simulated system. $R$ is the tip radius, $h$ and $L$ the thickness and the width of the simulated elastic medium, respectively. The medium is defined solely by its nominal Young's modulus $E$ and the Poisson's ratio $\nu$. As the only imposed boundary condition, the bottom face of the media is bonded to a rigid substrate. (**b**) Comparison of FCs simulated using FEA for the different systems and the different contact mechanics models (the theoretical curve for the TF-S system includes already the bottom effect through the corrective function $\Delta_D$).

### *3.1.1 Mesh Optimisation*

The proximity of the indenter-sample contact interface was finely meshed, and a coarser mesh was used moving away from it. Depending on indentation range, elements up to a minimum size of $0.026\ \mu m$ were used.

To optimise the mesh, we studied how the number of nodes affects the measured Young's modulus $E$ obtained through a Hertzian fit when a rigid sphere indents a sample sufficiently thick to avoid the bottom effect. Figure S6 shows that increasing the number of nodes above $N = 2 \cdot 10^4$ has a negligible impact on the value of the YM determined by extending the fit to increasingly wider ranges of indentation. Given that the computation time for carrying out a typical simulation on a desktop PC was below 60 minutes, we decided to use the maximum number of nodes allowed $N_{max} = 1.28 * 10^5$ within a given mesh configuration.

## 3.2. Data analysis

### *3.2.1. Hertz model*

Linearizing the experimental force vs indentation curves and the Hertz model (Eq. 1) would allow to apply a simple linear regression to the data to extract the Young's modulus. However, both measured force and indentation values $(F', \delta')$ are typically shifted with respect to the true values $(F, \delta)$ in Eq. 1. In terms of the measured quantities, the Hertz equation is:

$$F' - F_0 = \frac{4}{3}\frac{E}{1-\nu^2}\sqrt{R}(\delta' - \delta_0)^{3/2} \tag{13}$$

In the case of the force, as long as adhesion is negligible (in principle, the Hertz model assumes no adhesion at all; when measuring in liquid, adhesion is typically very small, as the Van der Waals interactions are effectively screened by water), the offset $F_0$ is due in general to a misalignment of the optical beam deflection system and can be easily subtracted during the pre-processing of the raw force curves. We will assume in the following that the offset on the force axis is not present, i.e. $F' = F$.

On the other hand, an apparent indentation $\delta' = \delta + \delta_0$ is computed when rescaling the distance axis in the raw force curve, where $\delta_0$ represents the unknown location of the contact point; $\delta_0$ must be therefore considered as a free parameter of the fit, as well as the Young's modulus. The presence of the offset $\delta_0$ in Eq. 13 does not allow to linearise the equation by taking the logarithm of both sides. Following Refs.[13,44], we adopt a convenient way to linearise Eq. 13 through the variable transformation $F^* = F^{2/3}$, which leads to the equation:

$$\delta' = \alpha F^* + \delta_0 \tag{14}$$

where the parameter $\alpha$ depends on the Young's modulus *E*:

$$\alpha = \left(\frac{3}{4}\frac{1-\nu^2}{E\sqrt{R}}\right)^{\frac{2}{3}} \tag{15}$$

and Eq. 14 holds for $\delta > \delta_0$.

The contact point $\delta_0$ represents the intercept of the $\delta'$ *vs* $F^*$ curve, while the Young's modulus $E$ can be calculated from the slope $\alpha$, since both radius $R$ and Poisson's ratio $\nu$ are supposed to be known. The free parameters of the fit, $\delta_0$ and $E$, are not correlated in Eq. 14, and can therefore be determined independently; this is not the case in the original Hertz model (Eq. 1).

To precisely identify the position where the probe gets in contact with the sample, we used the protocol described in detail by Puricelli et al.[13]. $\delta_0$ and $\alpha$ are left as free parameters while fitting the data. To find the contact point $\delta_0$ of each force curve, the force curves are rescaled, linearised and a linear fit is performed over the first 10% of the indentation range[13]: $\delta_0$ is the offset of indentation, which represents the point where a null force is exerted on the sample, thus it equals the point where the linear fit function intercepts the horizontal axis. The choice to fit the linearised Hertz model to the first 10% of the FC when a spherical indenter is used is supported by the observation that when $\gamma = \delta/R \leq 0.1$ (Figure 1d) the FCs simulated using parabolic and spherical indenters are in excellent agreement.

It is straightforward to generalise this approach to include large indentations and bottom effect corrections, or both. The correction functions $1/\Omega$ (Eqs. 6,7) and $1/\Delta$ (Eqs. 9a-12a and Eqs. 9b,12b), can be integrated into the pseudo force $F^*$ as follows[13]:

$$F^*_\Omega = \left(\frac{F}{\Omega}\right)^{2/3} \tag{16}$$

$$F^*_\Delta = \left(\frac{F}{\Delta}\right)^{2/3} \tag{17}$$

, where Eq.16 corrects for the large indentation effect, while Eq.17 corrects for the bottom effects.

In this work, we propose a way to combine both corrections into a single large indentation and bottom effect correction by defining the pseudo-force $F^*$ as:

$$F^*_{\Delta,\Omega} = \left(\frac{F}{\Delta \cdot \Omega}\right)^{2/3} \quad (18)$$

The resulting rescaled pseudo-Hertzian $F^*$ *vs* $\delta'$ curve (when both corrections are applied) is expected to be largely unaffected by both the large indentation issue, which depends on the specific geometry of the indenter (spherical instead of parabolic), and the finite thickness of the sample, and can then be fitted by the linearised Hertz model (Eqs 14,15) across the whole indentation range achieved experimentally, to retrieve the correct/intrinsic YM of the specimen. We demonstrate in this work that the correct YM can be retrieved with an accuracy better than 3%.

Eq.14 assumes that $F^*$ does not depend on $\delta_0$, which is true when no corrections are applied. However, the functions $\Delta$ and $\Omega$ depend on $\delta_0$ through the relation $\delta' = \delta + \delta_0$ (see Eqs. 16-17-18). In principle $\delta_0$ could be considered as a free parameter to be determined through a nonlinear fit, which would cause the loss of the advantages of the linearisation. In practice, one can follow the procedure describe above for the case when no corrections are applied to determine a first estimate of $\delta_0$ with good accuracy, then use Eqs. 14,16-18 with a new indentation axis shifted by $\delta_0$, indicating a residual intercept $\Delta\delta_0$ in place of $\delta_0$ in Eq. 14. Since $\Delta\delta_0$ is typically very small compared to the maximum indentation considered in the fit, neglecting it in the definition of the independent variable of the correction functions has no tangible effect.

### *3.2.2. Sneddon model*

If a spherical indenter is used and bottom effects are not important, the model developed by Sneddon can be used. This model accounts for arbitrarily large indentations, with the drawback of the lack of analytical solution. We assume here that both the measured force $F'$ and indentation $\delta'$ contain unknown offset: $F' = F + F_0$; $\delta' = \delta + \delta_0$. Eqs. 3,4 are then replaced by Eqs. 19,20:

$$F' - F_0 = \frac{E}{(1-\nu^2)}\left[(a^2 + R^2)\,\frac{(\delta\prime - \delta_0)}{a} - Ra\right] \quad (19)$$

$$\delta' - \delta_0 = \frac{1}{2}a \ln\left(\frac{R+a}{R-a}\right) \quad (20)$$

To obtain the YM from the experimental $F'$ vs $\delta'$ curve, the system of Eqs. 19,20 must be numerically solved, treating $E$, $\delta_0$ and $F_0$ as free parameters. This can be done by first solving numerically Eq. 20 using an initial guess for $\delta_0$, to obtain the relation $a(\delta')$, then substituting $a$ in Eq. 19 and evaluating the quadratic distance between the obtained curve and the experimental data; the best values of the free parameters $E$, $F_0$ and $\delta_0$ are those that minimise this distance. This method is time consuming, given that typically several hundreds of force curves must be processed.

### *3.2.3. Linearised Sneddon model*

We propose a faster method that once again takes advantage of linearisation. As for the Hertz model case, we assume that the vertical offset $F_0$ can be easily determined and subtracted. We then assume that the non-corrected, linearised Hertz model (Eq. 14) provides a rather accurate estimation of $\delta_0$ when applied to the first 10% of the indentation range, i.e. far from the large indentation limit. As for the force offset, also the indentation offset can be subtracted from the indentation axis, which allows to solve numerically only once the equation:

$$\delta = \frac{1}{2} a \ln\left(\frac{R+a}{R-a}\right) \tag{21}$$

, to obtain $a(\delta)$.

We then observe that the force in Eq. 19 depends linearly on the variable $s^*$ (with dimensions of an area) defined as:

$$s^* = \left[(a^2 + R^2)\ \frac{\delta}{a} - Ra\right] \tag{22}$$

, in terms of which Eq. 19 can be rewritten as:

$$F' = \frac{E}{(1-\nu^2)} s^* + \Delta F_0 \tag{23}$$

, where we have left for convenience, a residual force offset $\Delta F_0$. A linear regression of Eq. 23 provides the Young's modulus value.

While the standard approach, treating $E$, $\delta_0$ and $F_0$ as free parameters in Eqs. 19,20, requires a complete minimisation procedure for each FCs (the numerical inversion of Eq. 20 representing the bottleneck), the new approach requires the numerical inversion of Eq. 21, to be done only once, and then a series of linear regressions, which can be parallelise easily due to their algebraic nature, leading to a tremendous cut of computation time (the processing of a FV made of hundreds of FCs takes several minutes on a standard personal computer following the standard procedure and few seconds with the linearised one,). The free parameters $\Delta F_0$ and $E$ in Eq. 23 are independent, therefore a residual offset of the force does not affect the accuracy of the evaluation of the YM, assuming that $\delta_0$ can be determined with good accuracy.

### *3.2.4 Fitting the simulated force curves*

We simulated FCs with fixed values of both radius $R$ and thickness $h$ by varying the maximum indentation, corresponding to different maximum values of both $\gamma = \delta/R$ and $\chi = \sqrt{R\delta}/h$ parameters. Each single FC was simulated with 512 points and the fitted with the selected model up to 99% of their indentation. We therefore fitted the Hertz and corrected Hertz models to different (increasing) indentation ranges of the simulated FCs, and extracted the values of the Young's modulus, as described in the previous sections. On simulated bulk models, the maximum indentation was equal to the tip the radius ($\delta/R = 1$); in the thin film model this was not possible, since the convergence of the solution was not accurate, and we set the simulation with the maximum indentation $\delta/R = 0.6$.

## 3.3 AFM nanomechanical measurements

The nanomechanical measurements were performed in liquid (MilliQ water) using a Bioscope Catalyst AFM (Bruker) mounted on top of an inverted optical microscope (Olympus IX71). The whole system is isolated from the ambient noise by an active antivibration base (DVIA-T45, Daeil Systems) located inside an acoustic enclosure (Schaefer, Italy). We used custom colloidal probes, fabricated and calibrated as described in Ref. [14], produced attaching spherical borosilicate glass beads with different nominal radii R of 5 and 10 μm to tipless cantilevers (Nanosensors; TL-FM-50, TL-TM-50). The spring constants of the AFM probes, with nominal values of 3-5 N/m (FM) and 45 N/m (TM), were calibrated using the thermal noise method[45–47] and corrected for the contribution of the added mass of the sphere[48,49]. The radius of the fabricated colloidal probes after calibration turned out to be R = 4989 ± 35 nm, R = 7309 ± 13 nm, and R = 8898 ± 54 nm.

Samples were studied by collecting sets of deflection versus distance curves, also called force volumes (FVs), in different macroscopically separated locations. Each FV typically covered an area of 150x150 μm$^2$ and consisted of 15x15 or 20x20 curves (Figure S7a,b). In each experiment, the maximum applied force was set to achieve a maximum indentation $\delta \approx R$. Raw curves consist of the recording of deflection signal from the photodetector (in V units) vs the distance travelled by the z-piezo (in nm); these curves are rescaled into force vs indentation curves (shortly force curves, FCs) according to standard procedures, as described in Ref.[13] To this purpose, the deflection sensitivity of the optical beam deflection apparatus (in units of nm/V) was calculated using the contactless SNAP procedure[27], using the previously calibrated spring constant as reference.

From each FV, the local topographic map was reconstructed as the map of the contact points from each FC[50]. The mean-subtracted topographic map represents the map of local height variations around the mean thickness value of the sample in the region where the FV has been collected.

The Hertz model and the corrected models were fitted to each FC in the FV to extract the YM values. The median values of each FV were extracted by fitting a gaussian curve to the YM distributions in semi-$\log_{10}$ scale (Figure S7c, $E_{median}=10^{\log_{10}(E/Pa)|_{best}}$, where $\log_{10}()|_{best}$ is the centre of the gaussian curve), as described in detail in Ref[13].

## 3.4 Preparation of the reference elastic sample

Following Refs[51–53], samples of polyacrylamide (IUPAC poly(2-propenamide), abbreviated in PA) are prepared. Polyacrylamide is a polymer created from acrylamide (A) monomers that bind to each other via free radicals provided by an initiator forming monomeric chains that, in turn, bind to each other in a process driven by a cross linker (B) agent.

Polyacrylamide is a suitable material to study the elastic properties of thin films because of its relative ease of preparation, the possibility to tune its Young's modulus of elasticity through a suitable proportioning between monomer and cross-linker[51,52], and because it maintains a linear elastic behaviour up to large deformations[54–56] (see also Gavara et al[20]). The Young's modulus reflects the degree of structural organisation of the polymer at the level of molecular cross-linked chains, therefore changing the ratio monomer/cross-linker affects this organisation.

For the production of PA gels, we used acrylamide as the main component, bis-acrylamide (N,N*0*-Methylenebiscrylamide) as cross-linker and Irgacure 2959 as photoinitiator sensitive to UV light (wavelength 365 nm). Reagents were purchased from Sigma-Aldrich (St. Louis, MO). We prepared 50 mL of mother solution with concentrations of acrylamide $\rho_A = 0.12$ g/mL, *bis*-acrylamide $\rho_B = 0.0025$ g/mL and photoinitiator 0.001 g/mL as follows: first, the proper amounts of acrylamide and bis-acrylamide were added to 50 mL of Milli-Q water and stirred for 30 minutes; then, the photoinitiator was added, the solution was covered to protect it from environmental light and was stirred for another 30 minutes. The solution was then poured into a 3D-printed moulds of desired shapes (see Figure 2), which were composed of lateral walls with no upper and lower covers. The bottom of the mould consisted of a glass slide pre-treated with plasma oxygen (Diener), while the top cover was a glass slide not treated with plasma. To polymerise, the samples were exposed to a 365 nm UV source[53]. After the polymerisation, the mould and the upper glass were removed and the samples were left swelling in Milli-Q water overnight, to reach the thermal and chemical equilibrium. All the tested samples were obtained from the same mother solution and moulds to ensure reproducibility.

Two types of PA samples were prepared, based on their thickness $h$, measured as the height with respect to the rigid glass substrate: bulk samples ($h \sim 1\ mm$, Figure 2a), which are not affected by the bottom effect, and samples with a thickness gradient (Figure 2b), which can show the dependence of the measured YM on the local thickness.

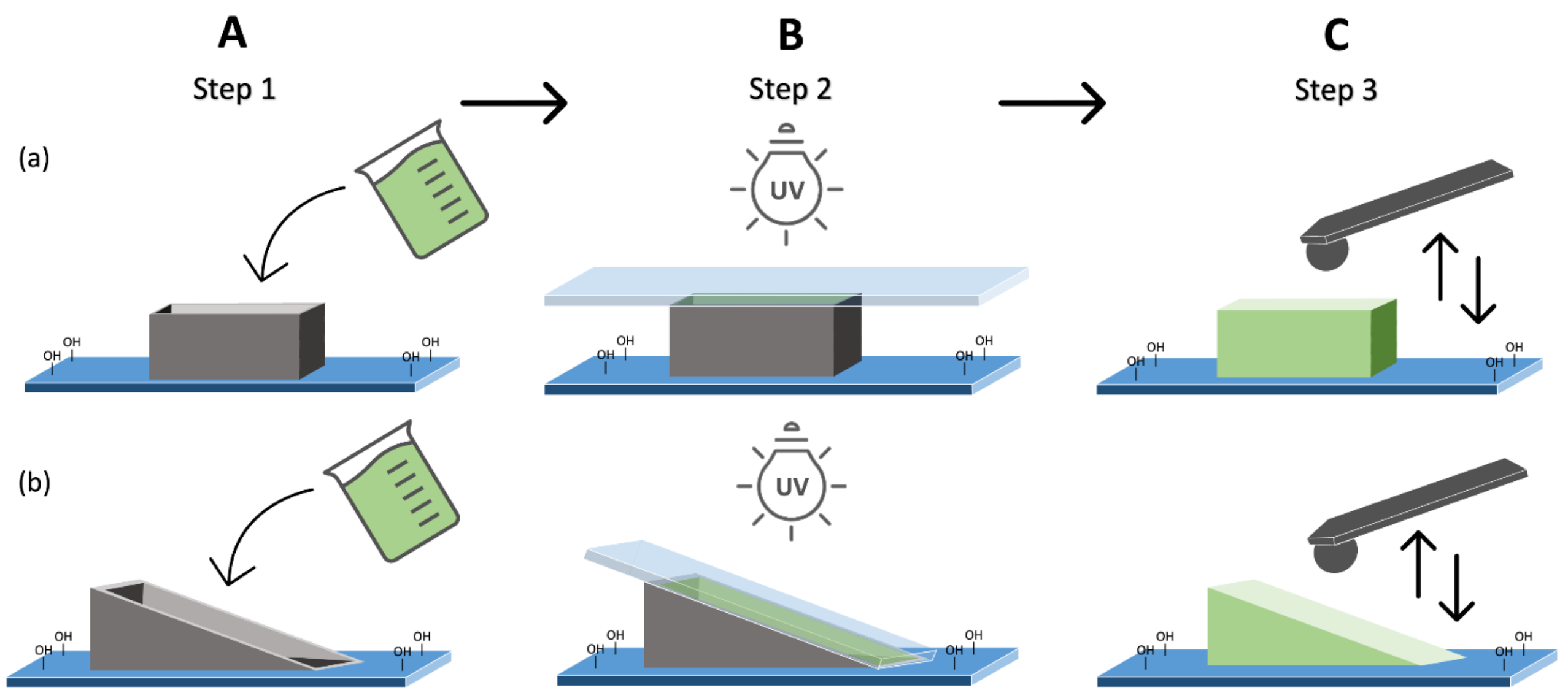

**Figure 2.** Representation of the main steps for the production of different elastic PA samples, prepared **following** closed moulding process. (**a**) *Bulk Sample*. A squared mould of height ≈ 1 mm closed by two glass slides is used. After the polymerisation of the poured solution, the upper coverslip and the mould are removed and then the gel was left to swell overnight in MilliQ water. The PA sample thickness is approximately 1 mm. (**b**) *Thickness-Gradient Sample.* A mould with variable height (maximum height 1 mm) and a 1 cm long edge is used (the inclination is 5.7°). After the polymerisation of the poured solution the upper coverslip and the mould are removed, and the swelling of the gel takes place. The sample possesses a linearly increasing thickness.

Knowledge of the local thickness of the sample is crucial to implement the BEC. We measured the local thickness of the PA sample (Figure 3a) by repeatedly engaging the AFM tip on the film and on the substrate and recording the distance Z travelled by z-microtranslator of the AFM head (used for the coarse approach of the tip to the sample), together with the XY coordinates of the precision motorised stage. The resolution of the stepper motors is as good as 100 nm, therefore allowing to reconstruct a “low-resolution” topographic map Z(X,Y) of the PA film (Figure 3b). The points measured on the glass slide were used to fit and subtract an eventual baseline of the low-resolution topographic map.

Having characterised the thickness gradient of the PA film, it was then possible to associate to each FV the mean thickness value of the measured portion of the sample by recording the XY position of each FV and looking at the corresponding Z value in the PA film topographic map (Figure 3b); the local height corresponding to each FC in the FV was then precisely determined by adding to the mean Z position of the FV the local height variation from the mean-subtracted, unflattened topographic map reconstructed from the FV (Figure 3c).

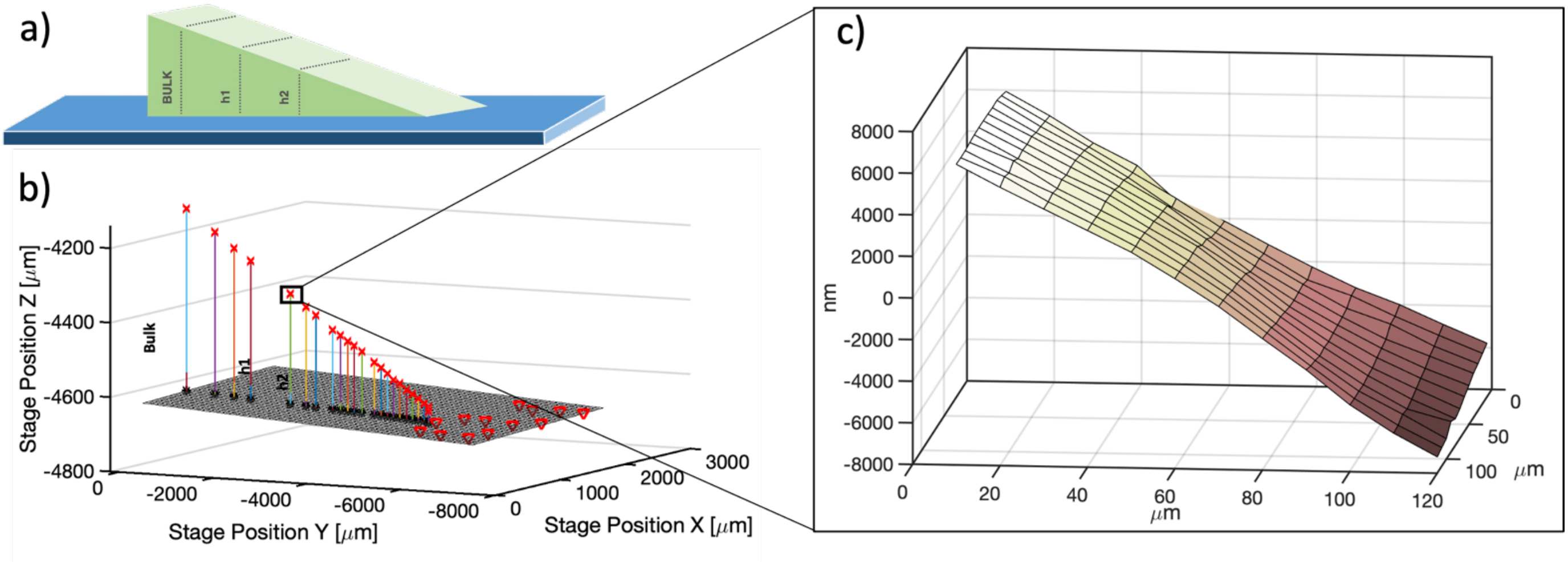

**Figure 3.** (**a**) Schematic representation of the samples possessing a gradient of thickness *h* prepared according to the closed moulding process. (**b**) Experimental reconstruction of the sample height profile based on the measurement of local Z(X, Y) coordinates, using the z-microtranslator of the AFM head and the precision motorised micro-translation XY stage of the AFM, as described in the main text. (**c**) 3D plot representing the map of the local height variations around the mean thickness value of the FV, corresponding to the mean-subtracted contact point map, obtained from the FCs., reconstructed from the FCs.

## 3.5 Error analysis in FEA

Since the Hertz model is mathematically exact for a paraboloidal indenter on a flat elastic medium for arbitrarily large indentations, we assume that the discrepancy between the nominal YM ($E_{\mathrm{nom}}$) and the one obtained fitting the Hertz model to the simulated force curve ($E_{\mathrm{Hertz}}$) depends only on simulation-related artifacts (such as the finite mesh size) and can therefore be assumed as the smallest error that can be obtained. The error associated to the fit of the Sneddon model for the spherical indenter to the simulated curves for arbitrary indentations can be estimated in a similar way.

Figure 4a shows that these errors are well below 1% across the widest range of indentations considered in this work ( $0 < \delta/R \leq 1$ ).

### 3.6 Error analysis in experimental measurements

Median YM values from different FVs corresponding to the same condition have been calculated and averaged, to obtain a mean median value; the corresponding error has been calculated as the standard deviation of the mean[13,57]. When multiple FV corresponds to the same condition (e.g. in the bulk samples, Figure 7a,b) the median values of each FV are collected into violin plots.

## 4. RESULTS

### 4.1. Validation of contact mechanics models and corrections by FEA

We have used FEA to study the range of applicability of the Hertz and Sneddon model, as well as bottom effect and large indentation corrections (the models used in the simulations are listed in Table 1). The Hertz model is exact for the paraboloidal indenter, while the Sneddon model is exact for the spherical indenter; both models, when used for the appropriate indenter geometry, are valid in principle for arbitrarily large indentations, if the material response is linearly elastic.

#### *4.1.1. Hertz and Sneddon models*

Artificial FCs produced simulating the B-S and B-P systems were fitted using the Hertz, Sneddon, and Sneddon linearised models up to increasing $\delta/R$ ratios. Figure 4a shows that on B-S systems the result of the fit performed using the Sneddon model agrees within 0.7% with the nominal Young's modulus; a slightly larger maximum discrepancy (1%) is observed when the B-P system is analysed using the simple Hertz model.

Noticeably, the linearised Sneddon (Figure 4a, purple crosses) model performs similarly, if not better, than the standard Sneddon model, across the whole range of indentations, which represents a validation of our linearised Sneddon approach. Eventually, our results confirm that when the appropriate contact mechanics model is applied with respect to the indenter geometry (i.e., Hertz model over a parabolic indenter and Sneddon model over a spherical indenter), at least for bulk films, the result is independent on the range of indentation selected for the fit, up to $\delta/R = 1$, as far as the system is linearly elastic.

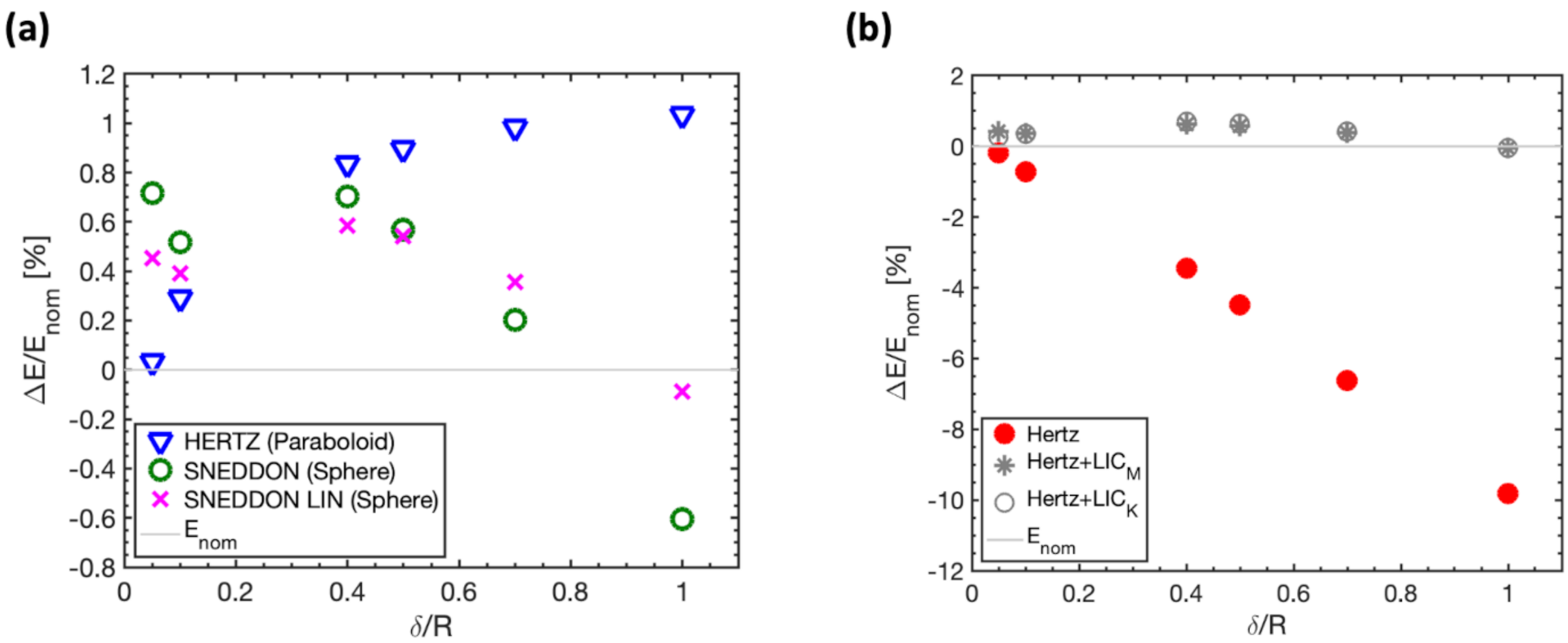

**Figure 4.** Relative discrepancies of the measured Young's modulus E compared to the nominal one ($E_{nom} = 0.5\ MPa$) for the bulk simulations for different ranges of indentation. The legends report the contact mechanics models used for the fit and the simulated systems (B-S and B-P) (see also Table 1). (**a**) Results for the appropriate matching of contact mechanics model and indenter geometry (Hertz model for the parabolic indenter, Sneddon and linearised Sneddon models for the spherical indenter). (**b**) Results for the Hertz model applied to the B-S system, with and without LIC.

### *4.1.2. Hertz model and large indentations*

The Hertz (Section 2.1) and Hertz+LIC (Section 2.3) models have been used to fit the FCs for the B-S system. The results are shown in Figure 4b. When the Hertz model is applied for indentations up to the tip radius, it underestimates the true Young's modulus of the media, with an error of approximately 10% when $\delta = R$; the error decreases almost linearly as the maximum indentation decreases, following the same trend of the LIC corrections (see Figure S2a). The Hertz model, which is exact for the paraboloidal indenter, does not accurately reproduce the indentation by a spherical indenter when the condition $\delta \ll R$ is not met. The Hertz+LIC models recover the true Young's modulus within less than 1% (by default) for all indentation ranges up to $\delta = R$ (the correction proposed by Muller provides slightly more accurate results).

### *4.1.3. The Hertz model and the simultaneous correction of bottom and large indentation effects*

The effect of the spatial (vertical) confinement of the sample was explored by fitting Hertz, Hertz+BEC and Hertz+BEC+LIC models to the FCs simulated on the TF-S systems (see Table 1). Different values of the maximum indentation $\delta$, sample thickness $h$ and tip radius $R$ were used to cover a wide range of $\chi$ values, up to $\chi = 0.4$. We have extended the simulation work of Garcia et al. [19] by considering different routes in the parameter space leading to the same maximum $\chi$:

variable indentation $\delta$ at fixed $R, h$; variable medium thickness $h$ at fixed $\delta, R$; variable tip radius $R$ at fixed $\delta, h$.

Figure 5 shows the relative discrepancies of the measured Young's modulus $E$ from the nominal one used in the simulations of the TF-S systems. Figure 5a,b,c show the results of the application of BEC and LIC separately, whereas Figure 5d,e,f show the results of the combination of LIC and BEC.

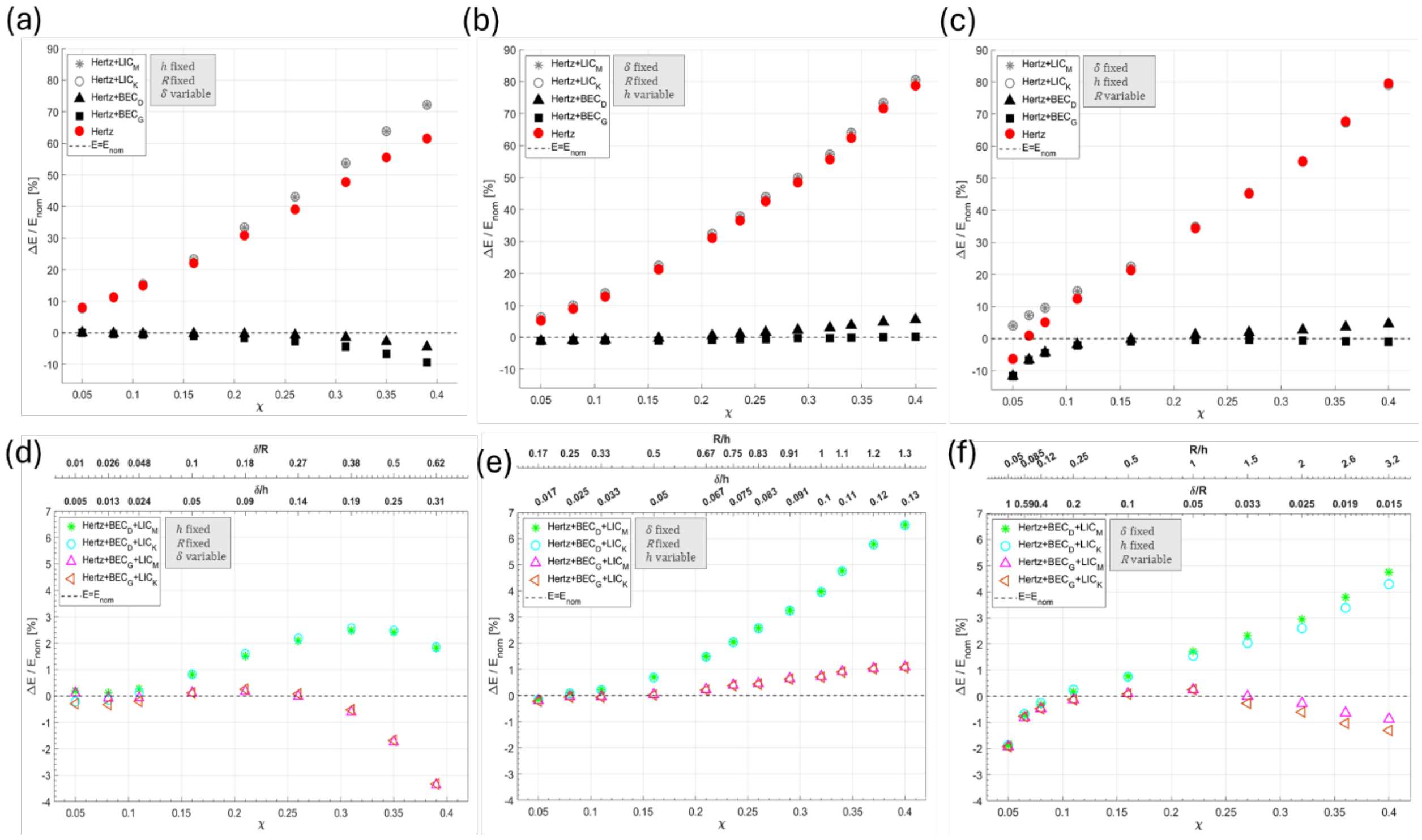

**Figure 5.** Relative discrepancies of the measured Young's modulus E compared to the nominal one ($E_{nom}$ = 0.5 MPa) for the TF-S systems for different maximum values of the parameter $\chi = \sqrt{R\delta}/h$. In the first row of plots (a-c), LIC and BEC were applied separately. In the second row (d-f), the combined correction LIC+BEC was applied (notice the dramatic reduction of the $\Delta E/E_{nom}$ range). (a,d) The maximum indentation was varied while fixing the medium thickness and the tip radius (third row of Table 1). (b,e) The medium thickness was varied while fixing the indentation and the tip radius (fourth row of Table 1). (c,f) The tip radius was varied while fixing the indentation and the medium thickness (fifth row of Table 1).

As long as one considers the effect of large indentation and finite thickness separately (Figure 5a,b,c), the same trend is observed: as $\chi$ increases, the relative discrepancy increases, in absolute terms. Fitting the Hertz model without considering any correction leads up to an 80% overestimation of the YM for $\chi = 0.4$ (red circles). Noticeably, the bottom effect is mostly responsible of this large discrepancy, the large indentation effect contributing for maximum 10% as $\delta/R$ approaches 1, confirming the results obtained on bulk systems (Figure 4b). The bottom effect dominates the inaccuracy of the Hertz model already at moderate values of $\chi$; a discrepancy of 5-8% is observed already at $\chi = 0.05$. We also notice (Figure 5a, red vs grey markers) that the large indentation effect leads to underestimation of the true YM, opposite to the bottom effect (Figure 5a, red vs black markers), which leads to a marked overestimation of the YM. As long as a maximum 10% underestimation of the YM is acceptable, the application of BEC only represents a valid solution.

When both BEC and LIC are applied, the maximum relative discrepancy of the measured YM from the nominal value can be further reduced to ±3% when Garcia's BEC is used (Figure 5d,e,f), in the interval $0 < \chi < 0.4$. When coupled to a BEC, Muller's and Kontomaris' large indentation corrections perform similarly; however, when $\delta/R$ is small (down to $\delta/R = 0.01$), Muller's LIC associated to Garcia's BEC leads to a smaller YM discrepancy (Figure 5d,f); indeed, in this small $\delta/R$ regime, Muller's correction is more accurate (see Figure S2a).

For similar maximum $\chi$ values, Dimitriadis' BEC may perform worse than Garcia's, in combination with a LIC (Figure 5e,f). This result may seem at odd with the fact that BECs depend only on $\chi$, therefore we could expect, for the same maximum $\chi$, a similar difference between the accuracy of the two BECs (i.e., a similar relative discrepancy of the YM); instead, Figure 5d,e,f show that, while Garcia's BEC overall performance is rather independent on which parameter is varied, Dimitriadis' is not. An explanation of the observed behaviour is that Garcia's BEC uses a more general expression for the contact radius, which takes explicitly into account the vertical confinement (Garcia's contact radius $a_G$ can be written as $a_G = a_H(1 + p_1\chi + \cdots)$, where $a_H$ is the Hertzian contact radius and $p_1$ is a constant, see Eq. S23 in Ref.[19] and Note S2), while Dimitriadis' BEC assumes a purely Hertzian contact radius. It follows that Garcia's BEC, through the $\chi$-dependent contact radius, can account for the different combinations of $R$, $h$, $\delta$ parameters, therefore providing a similar performance irrespective to how a specific maximum $\chi$ value is obtained; on the contrary, Dimitriadis's BEC, based on the Hertz radius, which is accurate only in the limit of the semi-infinite elastic layer, is more sensitive to specific combination of those parameters. Considering that $\chi = \sqrt{R\delta}/h = \sqrt{R/h \cdot \delta/h}$, we can expect a stronger residual dependence of the relative YM discrepancy on the two $R/h$ and $\delta/h$ ratios for Dimitriadis' BEC. Figure 5d,e,f show indeed that the difference between Dimitriadis' and Garcia's BECs is stronger when the $R/h$ is larger, while the $\delta/h$ ratio is far less impacting.

Our simulations confirm that due to their multiplicative nature, the large indentation and bottom effect correction functions can be applied together to the FCs obtained using a spherical indenter, to obtain sets of FCs that can be fitted using the standard Hertz model. Depending on the geometry of the system, the most accurate correction, according to our simulations, is generally obtained combining Garcia's BEC and Muller's LIC, though coupling Kontomaris' LIC provides only negligible worsening of the result in the limit of very small $\delta/R$ ratios. It is worth noting that, despite BEC only depends on the parameters $\chi = \sqrt{R\delta}/h$, the accuracy of the correction depends also on the way a specific maximum $\chi$ is obtained (i.e., on the specific combination of $R$, $h$, $\delta$).

## 4.2. Experimental validation of contact mechanics models and corrections

AFM nanoindentation experiments were performed on PA gels, both thick (bulk) and exhibiting a gradient of thickness (from the thin film to the bulk regime), produced according to the protocol described in Section 3.4. The contact mechanics models were fitted to the experimental curves up to the maximum indentation achieved. For the analysis of the experimental force curves, we selected Muller's and Garcia's models for LIC and BEC, respectively, since this combination performed better according to the FEA study presented in the previous sections.

### *4.2.1. Hertz and Sneddon models in the large indentation regime*

Figure 6 shows a representative set of FCs acquired using a colloidal probe with $R$ = 8.9 µm in a region of a PA gel with thickness of approximately 400 µm, which we can consider as bulk ($\chi$ = 0.018). Figure 6a shows the force vs indentation curves, linearised according to Eq. 14 (the original FCs are shown in the inset); a deviation from linearity of the non-corrected rescaled FCs is observed in Figure 6a for $\delta > 5$ µm , corresponding to $\delta/R \simeq 0.56$. Figure 6b shows the same FCs linearised and corrected for the large indentation effect, according to Eq. 16. The linearity of the rescaled, corrected FCs is remarkable, especially considering that this trend is conserved up to a maximum indentation of approximately 9 µm ($\delta/R \simeq 1$).

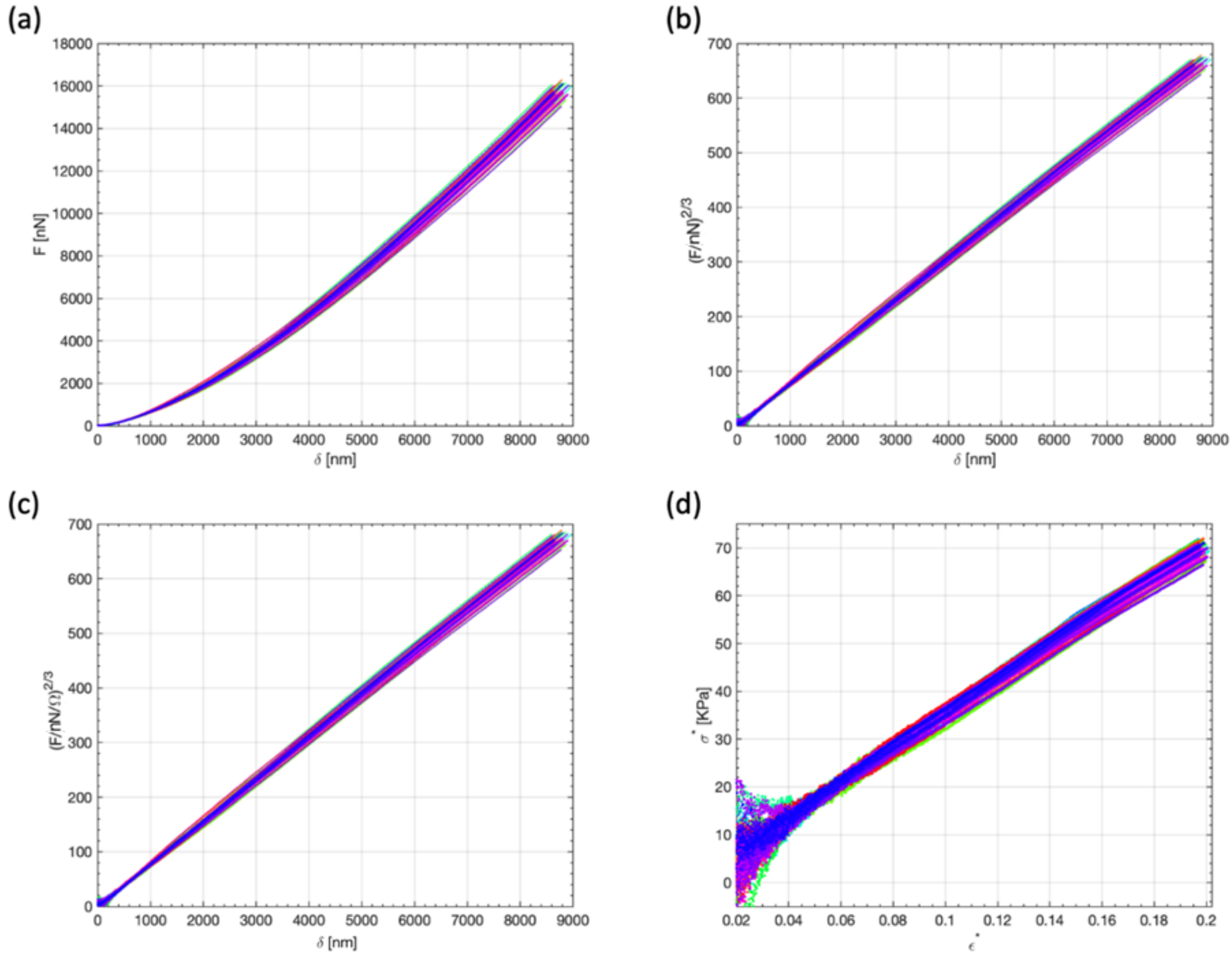

**Figure 6.** The force vs indentation curves acquired on a 700 μm thick PA gel, (**a**) Original FCs; (**b**) FCs rescaled according to Eq. 14.; (**c**) FCs rescaled and corrected according to Eq. 16 (LIC). In (**d**) the effective stress σ* vs strain ε* curve (Tabor; see Note S3) after the application of LIC is shown

That the linearity of the rescaled, corrected FCs shown in Figure 6b is a sign of a linear elastic behaviour of the PA sample, is further suggested by the observed scaling of the effective stress vs effective strain curves calculated using the Tabor's relations[30]: $\varepsilon^* = 0.2a/R$, $\sigma^* = F/(\pi a^2)$, after the application of LIC (Figure 6c). The stress-strain scaling for the PA film is linear up to a strain of 20%; the maximum stress is approximately 70 kPa, about half the Young's modulus of the PA gel, as measured in the bulk region. The result reported here is due to a combination of factors: the excellent linear elastic behaviour of PA gels and the use of large colloidal probes, which are effective in keeping stresses and strains low, distributing them across a large volume in the sample. Incidentally, the large deformation field is what makes the bottom effect important in nanomechanical measurements performed with colloidal probes. Sharp tips are far less affected by the bottom effect, though they can easily stimulate the nonlinear elastic response of the material.

Figure 7a shows the results of the experiments performed on two bulk samples made from the same working solution and mechanically tested using two different probes, one with $R \simeq 5.0$ μm and the other with $R \simeq 7.3$ μm, yet achieving the same maximum indentation ($\delta \simeq 4.2$ μm), corresponding to maximum $\delta/R$ values of 0.83 and 0.58, respectively. The Young's moduli of each bulk sample obtained using the different contact mechanics models (Sneddon, linearised Sneddon, Hertz and Hertz+LIC) are shown. For each sample, the linearised Sneddon model agrees with the Sneddon model with less than 1% discrepancy. The agreement of the corrected Hertz models with the Sneddon models is also excellent and within 1%. These experiments confirmed that the application of the Hertz model without LIC leads to the underestimation of the measured YM (see red violin plots in Figure 7a)

The agreement between Sneddon and linearised Sneddon models demonstrates that the assumption of negligible adhesion in experiments was reasonable; indeed, the proposed linearisation of the implicit Sneddon system works only under the hypothesis that any adhesion offset can be neglected. Using a linearised form of the Sneddon model represents a great advantage in data analysis. However, given that the results obtained by Hertz+LIC deviate from those of the Sneddon model by only 0.6%, it is clear that the corrected Hertz model represents an even better tool, due to its simplicity, for the fit of the force vs indentation curves, when spherical tips are used.

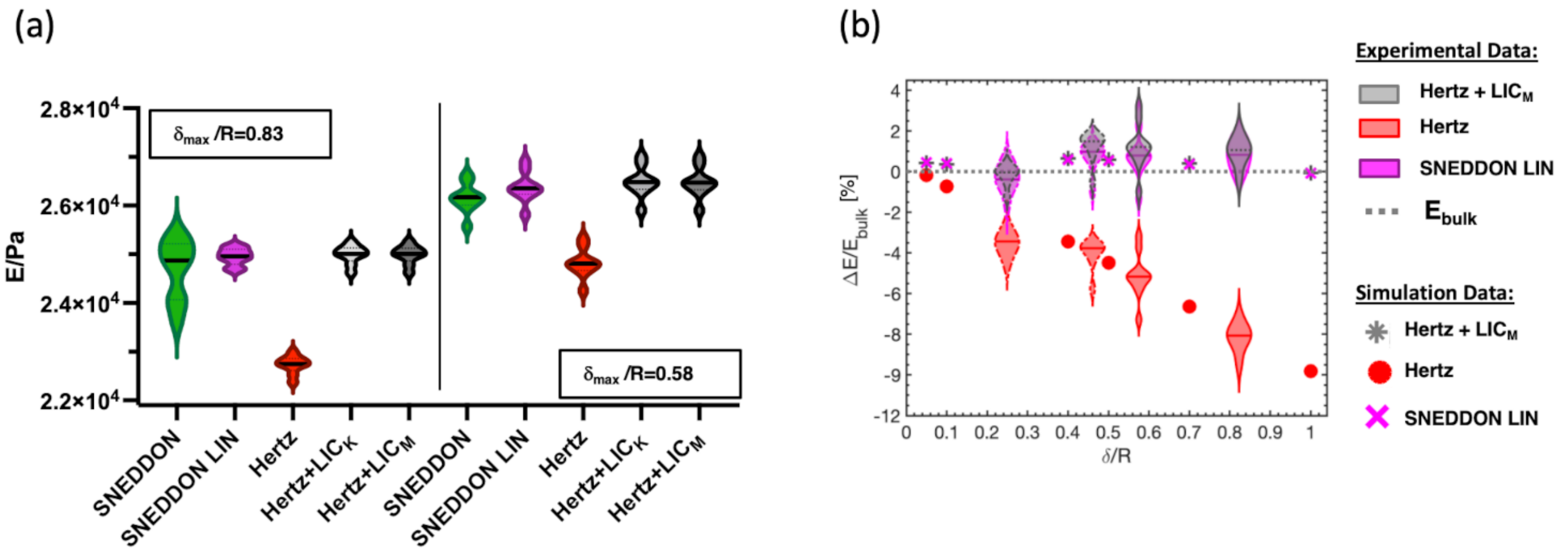

**Figure 7.** (**a**) Measured YM on two different bulk samples using two probes with different radii (the maximum indentation was the same, corresponding to maximum δ/R values of 0.83 and 0.58). The experimental FVs have been analysed using the five different contact mechanical models described in the Methods and reported in the label of the horizontal axis of the graph; each single point in a violin plot corresponds to the median YM value of a single FV, analysed with the described models. (**b**) Relative discrepancy between the YM measured using the different models and the value measured using the Sneddon model. The experimental results are shown as violin plots (single YM median values have been omitted for sake of clarity) together with the results of the simulations (single dots), described in the previous sections. The results obtained by applying the simple Hertz model without corrections are shown in red, while the results obtained applying the linearised Sneddon model are shown in purple. The grey violin plots represent the results obtained using the Hertz

model corrected with the Muller polynomial ($LIC_M$); the dashed violin-plots represent the results obtained by fitting the contact mechanics models to a smaller portion of the indentation axis.

To better highlight the large indentation artifacts and the efficacy of the correction procedure, the relative discrepancy between the Young's modulus measured using the Hertz-derived models and the one measured using the Sneddon model ($E_{\mathrm{bulk}}$) are shown in Figure 7b; results from both experiments and simulations are shown together (in the case of simulation, $E_{\mathrm{bulk}}$ corresponds to the nominal YM). Data are normalised with respect to the Young's modulus obtained through the Sneddon fit, because FEA showed that on bulk samples this model is very accurate, implicitly including the large indentation effects; it represents the experimental equivalent of the nominal YM in the simulations. Figure 7b confirms that the large indentation effect causes an underestimation of the true YM. When a spherical tip is used, neglecting the correction of the large indentation effect, i.e., applying the simple Hertz model, causes, in the worst case ($\delta/R = 1$ ) an underestimation of the true YM value as large as 10%. Furthermore, if a smaller portion of the same experimental indentation curves is fitted with the Hertz model (the dashed violin-plots in Figure 7b) the underestimation of the YM decreases in agreement with the predictions of finite elements analysis. The linearised Sneddon model for the spherical indenter and the Hertz+LIC provides the same level of accuracy of the standard Sneddon model for all $\delta/R$ ratios up to unity.

These results show that using spherical tips it is possible to exploit the whole indentation range (at least up to $\delta = \mathrm{R}$) for the fit, regardless of the tip radius, provided the correct model is used: Sneddon (also linearised) or Hertz+LIC. The use of parabolic tips in combination with the Hertz model would provide, in principle, no error at all, since for this geometry the Hertz model is exact. In practice, however, parabolic tips are not present on the market. Sharp pyramidal tips with blunted apexes[58] mimic the parabolic profile for small indentations; however, sharp tips would likely enhance nonlinear elastic effects already at relatively small $\delta/R$ values, which is supported by the evidence that experimentally measured YM values are typically largely overestimated[15]. The availability on the market of nearly paraboloidal tips with larger radii of curvature (> 100 nm) would allow in principle to apply the simple Hertz model with no need of LIC across a broad range of $\delta/R$ values.

### *4.2.2. The Hertz model and the simultaneous correction of bottom and large indentation effects*

Given the negligible differences between the results obtained using the Sneddon model and the Hertz+LIC model and considered the computational advantages of the linearised corrected Hertz models (as discussed in Sections 3.2.1), we have used the Hertz+LIC model to further investigate the BEC correction. Indeed, to our knowledge, models implementing BEC on the original implicit Sneddon system are not available.

To experimentally test the efficacy of combined BEC and LIC, we acquired several FVs on samples possessing a thickness gradient with a probe with radius $R$ = 8.9 μm and a maximum indentation of $\delta$ = [5-8] μm, corresponding to $\delta/R = 0.56 - 0.89,$ then we applied the simple Hertz model, Hertz+BEC, Hertz+LIC, and the two corrections coupled together, resulting in Hertz+ LIC+BEC.

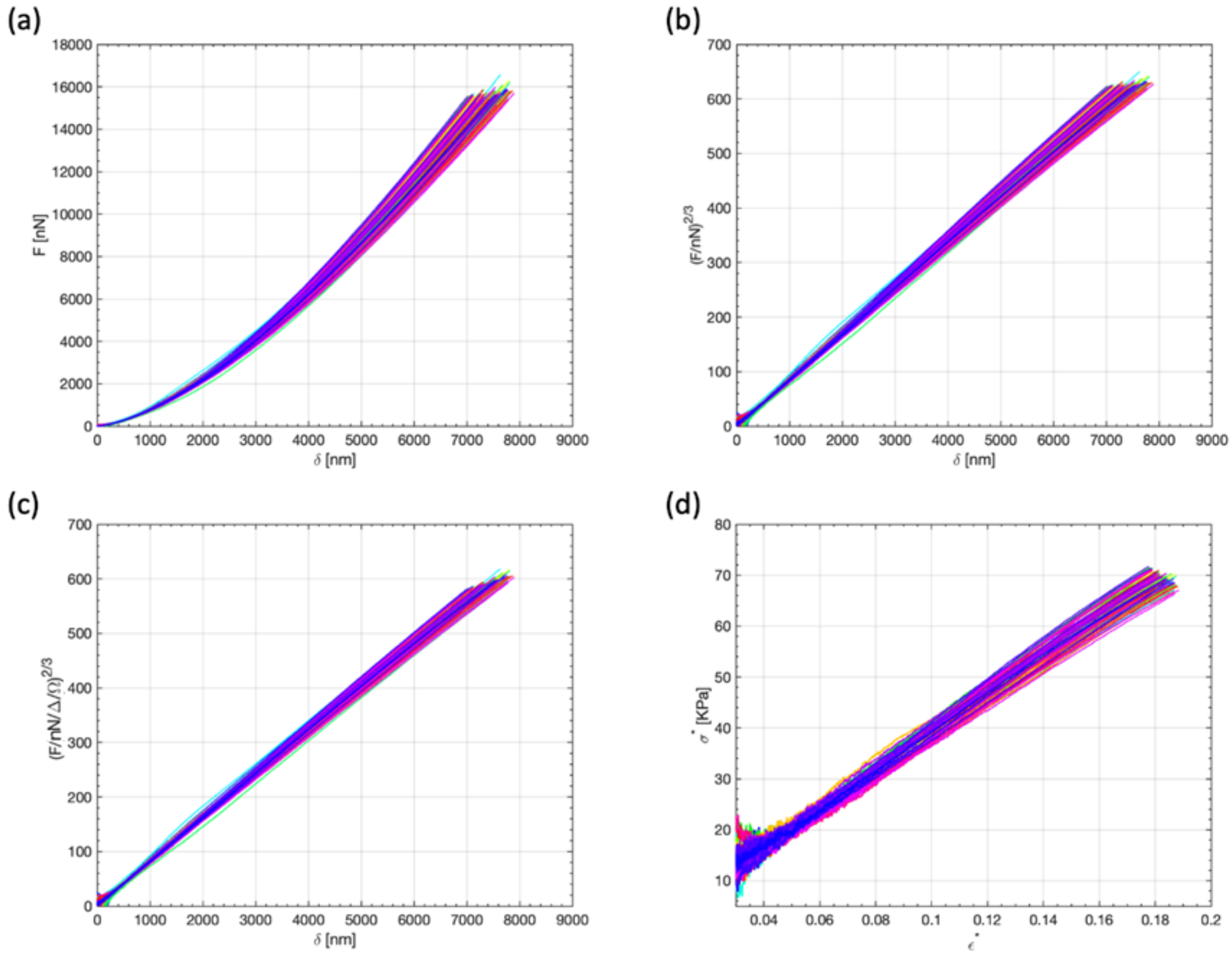

**Figure 8.** The force vs indentation curves acquired on a 47 μm thick PA gel. (**a**) Original FCs; (**b**) FCs rescaled according to Eq. 14.; (**c**) FCs rescaled and corrected according to Eq. 18 (LIC + BEC). In (**d**) the effective stress σ* vs strain ε* curve (Tabor; see Note S3) is shown.

Figure 8 shows a representative set of FCs acquired using a colloidal probe with $R$ = 8.9 μm in a thin region (approximately 47 μm) of a PA gel possessing a gradient of thickness. Figure 8a shows the force vs indentation curves, rescaled according to Eq. 14 (the original FCs are shown in the inset). No significant deviations from linearity are observed in the rescaled FCs acquired in the presence of a bottom effect, at odd with the case of the large indentation effect; the bottom effect mainly affects the overall slope of the rescaled force-indentation curve rather than distorting it. Figure 8b shows the same FCs rescaled and corrected for both large indentation and bottom effects, according to Eq. 18. The linearity of the rescaled, corrected FCs is very good, up to a maximum indentation of approximately 5 µm ($\delta/R \simeq 0.56$, $\chi = 0.13$). It can be noticed that the slope of the corrected linearised FCs is smaller than that of the non-corrected FCs, which is consistent with the expected stiffening of the film caused by the bottom effect.

Figure 8c shows that also in the case of the thin PA gel the effective stress vs effective strain curves, after the application of BEC and LIC, are linear up to the maximum induced strain of 15% and a maximum stress which is less than half the Young's modulus of the film; this confirms that also the PA sample with a gradient of thickness behaves like a linear elastic material.

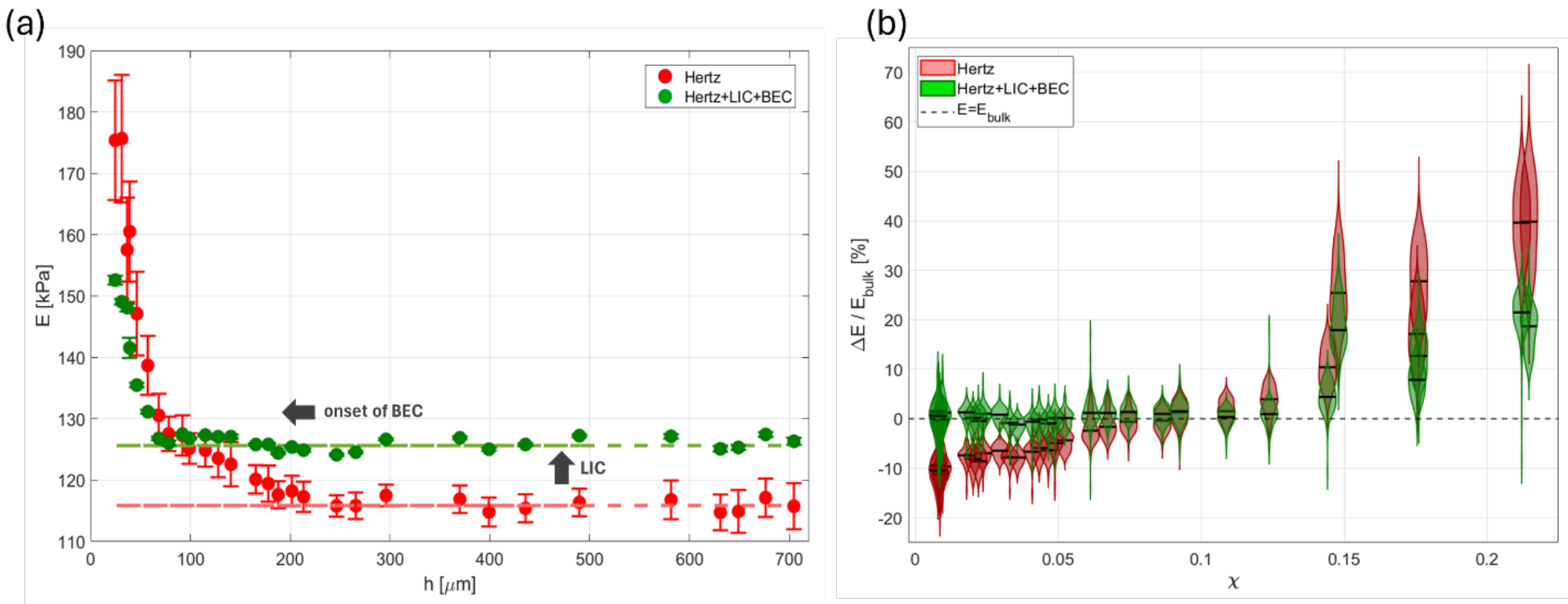

**Figure 9.** (a) The YM measured on PA samples with a thickness gradient, as a function of the local thickness. The results obtained fitting the Hertz and Hertz+LIC+BEC models, respectively, to the experimental data are shown. The red dashed line represents the apparent bulk Young's modulus, as measured by the Hertz model at large thicknesses. The green dashed line represents the corrected bulk Young's modulus, $E_{bulk}$, as measured by the Hertz+LIC+BEC model in the regions where the bottom effect is negligible ($\chi$<0.01). (b) The relative discrepancy between the measured YM and its bulk value as a function of the parameter $\chi$.

Figure 9a shows the YM vs thickness experimental curves, obtained using the Hertz and Hertz+LIC+BEC models. First, it can be noticed that the YM measured using the standard Hertz model has a marked dependence on the thickness (red symbols), for thickness up to 250 µm; for larger values of the thickness, the measured YM value tends to converge. However, the bulk YM value measured by the Hertz model is not accurate, since it is affected by the large indentation effect (at such large thicknesses, the bottom effect is negligible). The reference bulk YM value, $E_{\text{bulk}}$ (green line), has been calculated as the mean of the YM values obtained from those FVs where the differences between the values obtained using Hertz+LIC (not shown) and Hertz+LIC+BEC, respectively, is less than 1%; indeed, the two models are expected to provide the same correct bulk YM value in the bulk region. In agreement with the predictions of the LIC model and the results of FEA in the limit $\delta/R \rightarrow 1$, the difference between $E_{\text{bulk}}$ and the YM measured by the simple Hertz model, is approximately 10%.

Figure 9a shows that bottom effects are clearly impacting already at relatively small values of $\chi$ ($\chi > 0.03$ ), which for the $\delta/R$ ratios used in our experiments (large indentations and tip radius) corresponds to a thickness of approximately 200-250 µm. It is common habit to consider the bottom effect important when the $\delta/h$ ratio exceeds 0.1; our results confirm that the parameter $\chi = \sqrt{R\delta}/h$, rather than the $\delta/h$ ratio, should be considered to assess the opportunity of applying the bottom effect correction (especially when the tip radius is large). A partial mitigation of this caveat valid for large indentation experiments comes from the fact that in the intermediate $0.15 > \chi > 0.03$ region, bottom and large indentation effects are both present, with opposite effects.

Applying both LIC and BEC (green symbols) allowed us to obtain an experimental YM vs thickness curve that is flat across the widest range of thickness values, converging to the bulk YM value for large thicknesses. BEC very effectively removed the dependence of the YM on thickness in the 60-250 µm thickness range.

Plotting the relative discrepancy $\Delta E/E_{\text{bulk}} = (E - E_{\text{bulk}})/E_{\text{bulk}}$ against $\chi$ allows to better characterise the scaling of the measured YM and to compare experimental results with the predictions of the finite elements simulations. In Figure 9b the observed experimental discrepancies before and after the application of the Hertz+LIC+BEC model are shown. When both BEC and LIC are applied, the bulk YM is recovered within 1% up to $\chi = 0.13$ above which the discrepancy increases, for reasons that will be discussed later. The results of the nanomechanical measurements with both large indentation and bottom effect corrections applied on the PA gel are in very good agreement with those of FEA, following the same trend (for $\chi < 0.13$); while the discrepancy between the simulated and experimental results for the pure Hertz model is due to the different $\delta/R$ ratios obtained in simulations and experiments (LIC is not applied). This result and the evidence of the linear elastic response (Figures 6c and 8c) confirm the accuracy of the proposed combined correction of bottom and large indentation effects for the Hertz model when a spherical indenter is used.

For thickness below 60 µm, corresponding to $\chi > 0.13$ , the Hertz model corrections are not able to recover the bulk YM. At present we do not have a conclusive explanation for this unexpected behaviour, yet we can consider several evidences and make some hypothesis. Nonlinear elastic effects must be considered first, as potentially responsible for the residual nonlinearity observed in the corrected YM vs thickness curve. However, we tend to exclude that nonlinear elasticity plays a dominant role here because the Tabor's effective stress-strain curves provide clear evidence of linear elastic behaviour of our films, in both bulk and constrained configurations (Figure 6c, Figure 8c). This is consistent with the fact that PA is known as a highly linear elastic material. Moreover, the corrected rescaled FCs show excellent linearity across the entire range of indentation explored (Figure 6b, Figure 8b), as predicted by the Hertz model for a linear elastic material. We are prone to consider the unexpected stiffening of the PA gel as a real phenomenon; in fact, our presumed reference elastic PA samples are likely to possess a gradient of elasticity, and in particular are stiffer in the thinner regions.

Our failure in producing a reference elastic sample with uniform elastic properties, irrespective to the local thickness, could be due to the specific synthesis process that we used. The monomer and the crosslinker were mixed with a photoinitiator, and the solution was exposed to UV radiation for a relatively long time. Indeed, exposure time was increased to 30 minutes after we moved the UV lamp away from the sample to avoid overheating. It is not unlikely that a combination of nonuniform illumination and light absorbance across the sample thickness, consequent differential heating and diffusion of different species in the polymer solution may alter the crosslinking ratio and the material density in the thinner regions, increasing both, leading to a gradient of rigidity and a stiffening. Alternatively, chemical initiation of the crosslinking process could be exploited [51–53]. Thickness dependence of the Young's modulus of polymers has been observed and attributed to different organisation of polymeric chains during the synthesis process of thin films (for example as an effect of shear stress for spin-coated materials[59]). As the thickness of the substrate-bound film decreases, a strongly nonuniform, thickness-dependent swelling of the spatially constrained hydrogel (all measurements have been carried out in MilliQ water) can lead to reorganisation of the polymeric chains and to an internal pre-stress state (increasing with decreasing thickness), balanced by the surface tension of the PA gel, which could impact on the results of the nanomechanical measurements[60]. Moreover, the presence of a continuous gradient of thickness leads to an asymmetric redistribution of stresses inside the sample as the tip approaches the thin end. To mitigate the potential impact of nonuniform swelling and stress distribution inside the PA films, it could be possible to produce samples with a discrete, rather than continuous, variation of thickness, designing the mould to obtain flat terraces of different height rather than a uniform slope across the sample. Moreover, non-water-based, non-swelling elastic materials could represent a valid alternative to hydrogels, although potential issues related to higher Young's modulus values, viscoelasticity and elastic nonlinearity should be considered. Eventually, by increasing the radius of the tip and the maximum indentation, it would be possible to explore a similar range of $\gamma$ and $\chi$ values while measuring thicker regions, where the above-mentioned effects are expected to be less important.

The residual nonlinearity observed in our experiments deserves further investigations, which in turn stresses the importance of developing suitable test samples for the validation of mechanical models. It would be highly desirable to produce a highly elastic sample across a broad range of YM values [0.5-1000] kPa, with controlled nonlinearity. The lack of reliable, uniform and well characterizable elastic reference samples for nanomechanical analysis is at present still an open issue.

## 5. CONCLUSIONS

In this work, we have validated the application of Hertzian contact mechanics models and corrections in the framework of linear elasticity for the analysis of force vs indentation curves acquired using spherical probes by means of finite elements simulations and nanoindentation measurements on polyacrylamide gels with controlled thickness. We have systematically investigated the impact of both large indentations ($\delta/R \to 1$) and vertical spatial confinement ($\chi = \sqrt{R\delta}/h \to 0.4$) on the accuracy of the nanomechanical analysis performed with the Hertz model.

Our results demonstrate that, on bulk systems (thick samples, for which $\chi = \sqrt{R\delta}/h \ll 1$), the standard Hertz model can be used to fit force vs indentation curves acquired using colloidal probes up to indentation $\delta \leq 0.2R$ with an error below 2%, while both standard and linearised versions of the Sneddon model and the Hertz+LIC model can be used to fit force curves up to $\delta = R$ with approximately 1% accuracy. The linearised Sneddon and the Hertz + LIC models turned out to be as accurate as the standard Sneddon model, which is computationally more demanding.

We also demonstrated that the bottom effect leads to significant (> 5%) deviations from the bulk YM value for $\chi$ as small as 0.05, in the large indentation regime. We demonstrated that it is possible to combine BEC and LIC into a single correction function to apply the Hertz model to the FCs, greatly expanding its range of use and reliability, thus allowing to characterise samples possessing ample variations of thickness (height) on the scale of the tip dimensions. Such combined corrections are relevant when using micrometer-sized spherical tip in the large indentation regime (both $\delta/R$ and $\chi$ are large) and allow to take full advantage of the use of colloidal probes in the nanomechanical analysis.

The results of the nanoindentation measurements carried out on PA gels agree with the predictions of FEA and contribute to validate the proposed combined corrections of the Hertz model for force acquired using a spherical indenter. The agreement between FEA and experimental results is due to a combination of factors: the excellent linear elastic behaviour of PA gels and the use of large colloidal probes, which are effective in keeping stresses and strains low, distributing them across a large volume in the sample. Incidentally, the large deformation field produced by large spherical tips makes the bottom effect important in nanomechanical measurements performed with colloidal probes compared to sharp tips at similar indentations; however, sharp tips, besides possessing a not well characterizable geometry, can easily stimulate the nonlinear elastic response of the material. Test of the linearity of the elastic response should be always carried out, for example by looking at the scaling of the Tabor's effective stress vs strain curves.

The data analysis framework we presented, which is based on the linearisation of the Hertz equation, allows to include and test several different corrections among the one already published and those that will be published in the future.

The knowledge of the local sample height, or thickness, is crucial to implement the bottom effect correction. AFM offers the unique opportunity of obtaining both the topographic and the elastic maps from the same set of force curves; in those cases where the mean thickness of the sample cannot be obtained from the force curves (for example, because the underlying substrate is not accessible), an independent measure of the sample height must be obtained (for example by fluorescence confocal microscopy[28], profilometry, or exploiting the raw approach of the AFM head until engage, as described in this work).

The identification of reliable contact mechanics models and the standardisation of efficient and accurate data analysis methods are still open fields of investigation within the AFM and the nanoindentation community. Given the typical dimensions of the systems and indentation parameters, both large indentation and bottom effects have a strong impact on the nanomechanical characterisation of living cells as well as thin tissue and ECM slices and should be routinely implemented in the experimental and data analysis procedures. Overall, the accurate correction of both large indentation and bottom effects can reduce the variability between results coming from different laboratories and increase the reproducibility of the nanomechanical experiments, supporting the standardisation effort.

Moreover, our study demonstrated that an essential part of this standardisation effort must be the development of reliable, uniform and well characterizable elastic reference samples with negligible elastic nonlinearity and controlled YM in physiologically relevant range [0.5-1000] kPa.

**Author contributions**

Conceptualisation: MC, AP;

methodology – probe fabrication and characterisation: HH, MC, LDF;

methodology – Finite Element Simulations: LDF;

data curation and analysis HH, MC, LDF;

original draft writing and editing: MC, LDF;

draft revision: all authors;

supervision: MC, AP;

resources, funding, and project administration: AP.

Authors' contributions were allocated adopting the terminology of CRediT Contributor Roles Taxonomy.

**Conflicts of interest**

There are no conflicts to declare.

***Acknowledgements***

This research was partially funded by the European Union Horizon 2020 research and innovation program under the Marie Skłodowska-Curie grant agreement No. 812772, project Phys2BioMed. We acknowledge financial support under the National Recovery and Resilience Plan (NRRP), Mission 4, Component 2, Investment 1.1, Call for tender No. 104 published on 2.2.2022 by the Italian Ministry of University and Research (MUR), funded by the European Union – NextGenerationEU– Project Title Impact of chromatin organization on nuclear stiffness and cell migration – CUP G53D23002540006 - Grant Assignment Decree No. 2022KRWA7Y_002 adopted on 15-06-2023 by the Italian Ministry of Ministry of University and Research (MUR). We thank Andrei Shvarts and Lukasz Kaczmarczyk from the University of Glasgow for support and discussions on Finite Element Simulations; we thank Ricardo Garcia from the Institute of Science and Materials of Madrid, Andreas Stylianou from the European University of Cyprus and Stylianos Vasileios Kontomaris from the Metropolitan College of Athens for useful discussions.

# REFERENCES

[1] K. Takahashi, Y. Kakimoto, K. Toda, K. Naruse, *J Cell Mol Med* **2013**, *17*, 225.

[2] M. L. Jackson, A. R. Bond, S. J. George, *Cardiovasc Drugs Ther* **2023**, *37*, 997.

[3] S. W. Verbruggen, L. M. McNamara, in *Mechanobiology in Health and Disease*, Elsevier, **2018**, pp. 157–214.

[4] D. Ingber, *Ann Med* **2003**, *35*, 564.

[5] M. Lekka, D. Navajas, M. Radmacher, A. Podestà, Eds. , *Mechanics of Cells and Tissues in Diseases*, De Gruyter, **2023**.

[6] M. Chighizola, T. Dini, C. Lenardi, P. Milani, A. Podestà, C. Schulte, *Biophys Rev* **2019**, *11*, 701.

[7] H. Liu, L. A. MacQueen, J. F. Usprech, H. Maleki, K. L. Sider, M. G. Doyle, Y. Sun, C. A. Simmons, *Biomaterials* **2018**, *172*, 30.

[8] F. Iberite, M. Piazzoni, D. Guarnera, F. Iacoponi, S. Locarno, L. Vannozzi, G. Bolchi, F. Boselli, I. Gerges, C. Lenardi, L. Ricotti, *ACS Appl Bio Mater* **2023**, *6*, 2712.

[9] Y. F. Dufrêne, T. Ando, R. Garcia, D. Alsteens, D. Martinez-Martin, A. Engel, C. Gerber, D. J. Müller, *Nat Nanotechnol* **2017**, *12*, 295.

[10] D. J. MÜLLER, Y. F. DUFRÊNE, in *Nanoscience and Technology*, Co-Published With Macmillan Publishers Ltd, UK, **2009**, pp. 269–277.

[11] H. Holuigue, E. Lorenc, M. Chighizola, C. Schulte, L. Varinelli, M. Deraco, M. Guaglio, M. Gariboldi, A. Podestà, *Sensors* **2022**, *22*, 2197.

[12] E. Lorenc, H. Holuigue, F. Rico, A. Podestà, in *Biomedical Methods*, De Gruyter, **2023**, pp. 87–104.

[13] L. Puricelli, M. Galluzzi, C. Schulte, A. Podestà, P. Milani, *Review of Scientific Instruments* **2015**, *86*, 33705.

[14] M. Indrieri, A. Podestà, G. Bongiorno, D. Marchesi, P. Milani, *Review of Scientific Instruments* **2011**, *82*, 023708.

[15] A. Kubiak, M. Chighizola, C. Schulte, N. Bryniarska, J. Wesolowska, M. Pudelek, M. Lasota, D. Ryszawy, A. Basta-Kaim, P. Laidler, A. Podestà, M. Lekka, *Nanoscale* **2021**, *13*, 6212.

[16] H. Hertz, *Journal fur die Reine und Angewandte Mathematik* **1882**, *1882*, 156.

[17] L. Lacaria, A. Podestà, M. Radmacher, F. Rico, in *Biomedical Methods*, De Gruyter, **2023**, pp. 21–64.

[18] E. K. Dimitriadis, F. Horkay, J. Maresca, B. Kachar, R. S. Chadwick, *Biophys J* **2002**, *82*, 2798.

[19] P. D. Garcia, R. Garcia, *Biophys J* **2018**, *114*, 2923.

[20] N. Gavara, R. S. Chadwick, *Nat Nanotechnol* **2012**, *7*, 733.

[21] P. D. Garcia, C. R. Guerrero, R. Garcia, *Nanoscale* **2020**, *12*, 9133.

[22] S. V Kontomaris, A. Malamou, *Eur J Phys* **2021**, *42*, 025010.

[23] P. Müller, S. Abuhattum, S. Möllmert, E. Ulbricht, A. V. Taubenberger, J. Guck, *BMC Bioinformatics* **2019**, *20*, DOI 10.1186/s12859-019-3010-3.

[24] Á. dos Santos, A. W. Cook, R. E. Gough, M. Schilling, N. A. Olszok, I. Brown, L. Wang, J. Aaron, M. L. Martin-Fernandez, F. Rehfeldt, C. P. Toseland, *Nucleic Acids Res* **2021**, *49*, 340.

[25] I. N. Sneddon, *Proceedings of the Glasgow Mathematical Association* **1965**, *7*, 48.

[26] R. Long, M. S. Hall, M. Wu, C.-Y. Hui, *Biophys J* **2011**, *101*, 643.

[27] H. Schillers, C. Rianna, J. Schäpe, T. Luque, H. Doschke, M. Wälte, J. J. Uriarte, N. Campillo, G. P. A. Michanetzis, J. Bobrowska, A. Dumitru, E. T. Herruzo, S. Bovio, P. Parot, M. Galluzzi, A. Podestà, L. Puricelli, S. Scheuring, Y. Missirlis, R. Garcia, M. Odorico, J.-M. Teulon, F. Lafont, M. Lekka, F. Rico, A. Rigato, J.-L. Pellequer, H. Oberleithner, D. Navajas, M. Radmacher, *Sci Rep* **2017**, *7*, 5117.

[28] S. Pérez-Domínguez, S. G. Kulkarni, J. Pabijan, K. Gnanachandran, H. Holuigue, M. Eroles, E. Lorenc, M. Berardi, N. Antonovaite, M. L. Marini, J. Lopez Alonso, L. Redonto-Morata, V. Dupres, S. Janel, S. Acharya, J. Otero, D. Navajas, K. Bielawski, H. Schillers, F. Lafont, F. Rico, A. Podestà, M. Radmacher, M. Lekka, *Nanoscale* **2023**, *15*, 16371.

[29] C. Valero, B. Navarro, D. Navajas, J. M. García-Aznar, *J Mech Behav Biomed Mater* **2016**, *62*, 222.

[30] D. C. Lin, D. I. Shreiber, E. K. Dimitriadis, F. Horkay, *Biomech Model Mechanobiol* **2009**, *8*, 345.

[31] C. E. Wu, K. H. Lin, J. Y. Juang, *Tribol Int* **2016**, *97*, 71.

[32] K. D. Costa, F. C. P. Yin, *J Biomech Eng* **1999**, *121*, 462.

[33] S. V. Kontomaris, A. Stylianou, A. Georgakopoulos, A. Malamou, *Micron* **2023**, *164*, 103384.

[34] S. V. Kontomaris, A. Malamou, *Mater Res Express* **2020**, *7*, 033001.

[35] S.-V. Kontomaris, A. Malamou, *Eur J Phys* **2022**, *43*, 015010.

[36] A. C. Fischer-Cripps, *The Hertzian Contact Surface*, **1999**.

[37] I. N. Sneddon, *Int J Eng Sci* **1965**, *3*, 47.

[38] K. B. Park, M. S. Kim, J. H. Kim, S. K. Kim, J. M. Lee, *Journal of Polymer Engineering* **2019**, *39*, 432.

[39] J. R. Barber, K. A. Publisher, S. P. Timoshenko, H. Love, L. D. Landau, E. M. Lifshitz, **2007**, 1.

[40] S. R. Kalidindi, S. Pathak, *Acta Mater* **2008**, *56*, 3523.

[41] D. K. Patel, S. R. Kalidindi, *Acta Mater* **2016**, *112*, 295.

[42] B. R. Donohue, A. Ambrus, S. R. Kalidindi, *Acta Mater* **2012**, *60*, 3943.

[43] C.-E. Wu, K.-H. Lin, J.-Y. Juang, *Tribol Int* **2016**, *97*, 71.

[44] P. Carl, H. Schillers, *Pflugers Arch* **2008**, *457*, 551.

[45] H. J. Butt, M. Jaschke, *Nanotechnology* **1995**, *6*, 1.

[46] H. J. Butt, B. Cappella, M. Kappl, *Surf Sci Rep* **2005**, *59*, 1.

[47] M. Chighizola, J. Rodriguez-Ramos, F. Rico, M. Radmacher, A. Podestà, in *Biomedical Methods* (Eds: M. Lekka, D. Navajas, M. Radmacher, A. Podestà), De Gruyter, **2023**, pp. 105–128.

[48] M. Chighizola, L. Puricelli, L. Bellon, A. Podestà, *Journal of Molecular Recognition* **2021**, *34*, e2879.

[49] J. Laurent, A. Steinberger, L. Bellon, *Nanotechnology* **2013**, *24*, 225504.

[50] J. Domke, M. Radmacher, *Langmuir* **1998**, *14*, 3320.

[51] J. R. Tse, A. J. Engler, *Curr Protoc Cell Biol* **2010**, *47*, DOI 10.1002/0471143030.cb1016s47.

[52] A. K. Denisin, B. L. Pruitt, *ACS Appl Mater Interfaces* **2016**, *8*, 21893.

[53] S. Sheth, E. Jain, A. Karadaghy, S. Syed, H. Stevenson, S. P. Zustiak, *Int J Polym Sci* **2017**, *2017*, DOI 10.1155/2017/5147482.

[54] Z. Cao, Z. Yuan, R. Wu, H. Wu, B. Jin, J. Zheng, J. Wu, *Gels* **2021**, *7*, 177.

[55] C. Storm, J. J. Pastore, F. C. MacKintosh, T. C. Lubensky, P. A. Janmey, *Nature* **2005**, *435*, 191.

[56] A. Engler, L. Bacakova, C. Newman, A. Hategan, M. Griffin, D. Discher, *Biophys J* **2004**, *86*, 617.

[57] E. Lorenc, L. Varinelli, M. Chighizola, S. Brich, F. Pisati, M. Guaglio, D. Baratti, M. Deraco, M. Gariboldi, A. Podestà, *Sci Rep* **2023**, *13*, 12175.

[58] F. Rico, P. Roca-Cusachs, N. Gavara, R. Farré, M. Rotger, D. Navajas, *Phys Rev E Stat Nonlin Soft Matter Phys* **2005**, *72*, 1.

[59] M. Liu, J. Sun, Y. Sun, C. Bock, Q. Chen, *Journal of Micromechanics and Microengineering* **2009**, *19*, DOI 10.1088/0960-1317/19/3/035028.

[60] J. M. Long, G. F. Wang, *Mechanics of Materials* **2013**, *56*, 65.

Validation of contact mechanics models for Atomic Force Microscopy via Finite Elements Analysis and nanoindentation experiments

L. Dal Fabbro, H. Holuigue[1], M. Chighizola[2,]* and A. Podestà*

Dipartimento di Fisica "Aldo Pontremoli", Università degli Studi di Milano, via G. Celoria 16, 20133, Milano, Italy.

CIMaINa, Università degli Studi di Milano, via G. Celoria 16, 20133, Milano, Italy.

*Corresponding authors. E-mail: matteo.chighizola@ifom.eu, alessandro.podesta@unimi.it

# SUPPLEMENTARY INFORMATION

**Table of Contents**

[1] Present affiliation: Mechanobiology Institute (MBI), National University of Singapore, 5A Engineering Drive 1, 117411, Singapore

[2] Present affiliation: Institute FIRC of Molecular Oncology (IFOM), Via Adamello 16, 20139 Milan, Italy.

## Note S1. Contact radius of a spherical tip affected by large indentations

In the work of Kontomaris *et. al.*[1] it is explicitly reported an expression for the corrected contact radius $a$ when the spherical tip of radius $R$ can no longer be approximated by a paraboloid:

$$\frac{a}{R} = c_1\gamma^{1/2} + c_2\gamma + c_3\gamma^2 + c_4\gamma^3 + c_5\gamma^4 + c_6\gamma^5 \tag{S0}$$

, being $\gamma = \delta/R$ . The coefficients $c_k$ are reported in the main text (see Eq. 6).

On the other hand, Muller *et. al.*[2] do not present any formula for the contact radius in the p. We therefore explicitly computed the Muller's contact radius on the basis of his large indentations corrective factor $\Omega_M(\gamma)$.

We took advantage of the general relationship linking the contact radius $a$, the deformation $\delta$ and the applied force $F$ by an axisymmetric body indenting an elastic half space with Young's modulus $E$ and Poisson's ratio $\nu$:[3]

$$\frac{dF}{d\delta} = \frac{2E}{1-\nu^2}a \tag{S1}$$

When the large indentations effect is present, the applied force $F$ can be written as

$$F = F_H(\delta) \cdot \Omega_M(\delta/R) \tag{S2}$$

, being $F_H(\delta)$ the equivalent Hertzian force, valid for a paraboloidal indenter.

Inserting Eq. S2 into Eq. S1 and noting that $d\Omega_M/d\gamma \approx -0.1$ when $\gamma \in [0,1]$, it follows:

$$a \approx \Omega_M a_H - 0.1\frac{2}{3}\frac{\delta^{3/2}}{\sqrt{R}} \approx a_H(\Omega_M - 0.07\gamma) \tag{S3}$$

, where $a_H = \sqrt{R\delta}$ is the Hertzian contact radius. The corrective factor within round brackets in Eq. S3 depends only on the nondimensional ratio $\gamma$.

# Note S2. Garcia's bottom effect correction for an arbitrary Poisson's ratio and the formulae for the contact radius accounting for the bottom effect

### *Garcia's bottom effect correction for an arbitrary Poisson's ratio*

Within the framework of linear elasticity proposed by Garcia *et al.,* we solved Eq. S21 in the Supporting Information of Ref.[4], using a custom routine based on the Symbolic Math Toolbox of MATLAB (Mathworks). We obtained the expression of the corrective factor $\Delta_G$ for the bonded case for an arbitrary value of the Poisson's ratio up to the desired order in $\chi$. The corrective factor of Garcia $\Delta_G$, up to the fifth order, is:

$$\Delta_G = 1 + A\chi + B\chi^2 + C\chi^3 + D\chi^4 + E\chi^5 \quad \text{(S4)}$$

, where the coefficients are:

$$A = \frac{2\alpha_0}{\pi}$$

$$B \approx \frac{301\pi\alpha_0^2}{2000}$$

$$C \approx -\pi\left(\frac{31\alpha_0^3}{255} + \frac{106\beta_0}{491}\right)$$

$$D \approx \pi\left(\frac{24\alpha_0^4}{245} + \frac{40\beta_0\alpha_0}{97}\right)$$

$$E \approx -\pi\left(\frac{\alpha_0^5}{22} + \frac{2\beta_0\alpha_0^2}{9}\right)$$

, with:

$$\alpha_0 = -\frac{1.2876 - 1.4678\nu + 1.3442\nu^2}{1-\nu}$$

$$\beta_0 = \frac{0.6387 - 1.0277\nu + 1.5164\nu^2}{1-\nu}$$

The quantities $\alpha_0$ and $\beta_0$ are defined as in Dimitriadis *et al*.[5] for the bonded case. The given expression of the coefficient $A$ is analytically exact, whereas the ones of $B, C, D$ and $E$ well approximate the true values up to the 4th decimal point. The fifth-order term in Eq. S4 does not contribute significantly to the evaluation of the YM (less than 0.3%), hence we have used the expression of $\Delta_G$ truncated to the fourth order (Eq. 12b in the main text).

### *Garcia's contact radius expansion up to the 5th order in $\chi$*

In Garcia *et al.* the contact radius is evaluated as (Eq, S23 in the Supporting Information of Ref. [4]):

$$a_G = \sqrt{\delta R} - \frac{2\alpha_0 R\delta}{3h\pi} = a_H(1 + p_1\chi) \quad \text{(S5)}$$

, obtained through the minimization of the derivative of the first two terms of the force with respect to the contact radius (see Eq. S21-S22 in the SI of Ref. [4]); $a_H$ is the Hertzian contact radius, $\chi = \sqrt{R\delta}/h$ and $p_1 = -(2\alpha_0/3\pi)$. The corrective factor within round brackets in Eq. S5 depends only on the nondimensional ratio $\chi$.

In this work, we evaluated the expression of the contact radius within the framework of Garcia by performing the minimization of the derivative of the complete force expression provided in Eq. S21 of Ref.[4]. We found that also when considering the full force expression, the contact radius may be expressed as the Hertzian contact radius multiplied by a polynomial expansion in $\chi$. The expansion of the Garcia's contact radius (bonded case) up to the 5th order, for an arbitrary Poisson's ratio $\nu$, is:

$$a_G^{\text{full_exp}} = a_H(1 + p_1\chi + p_2\chi^2 + p_3\chi^3 + p_4\chi^4 + p_5\chi^5) \quad \text{(S6)}$$

, with coefficients $p_i$ depending solely upon $\alpha_0$ and $\beta_0$ :

$$p_1 = -\frac{2\alpha_0}{3\pi} = -0.2122\ \alpha_0$$

$$p_2 = 0.1126\ \alpha_0^2$$

$$p_3 = -\ (0.0765\ \alpha_0^3 + 0.5093\ \beta_0)$$

$$p_4 = 0.0585\ \alpha_0^4 + 0.5764\ \alpha_0\beta_0$$

$$p_5 = 0.1261\ \alpha_0^5 + 0.9250\ \alpha_0^2\beta_0$$

The coefficients were determined using a custom routine based on the Symbolic Math Toolbox of MATLAB (Mathworks). Also in this case, the corrective factor in Eq. S6 depends only on the nondimensional ratio $\chi$.

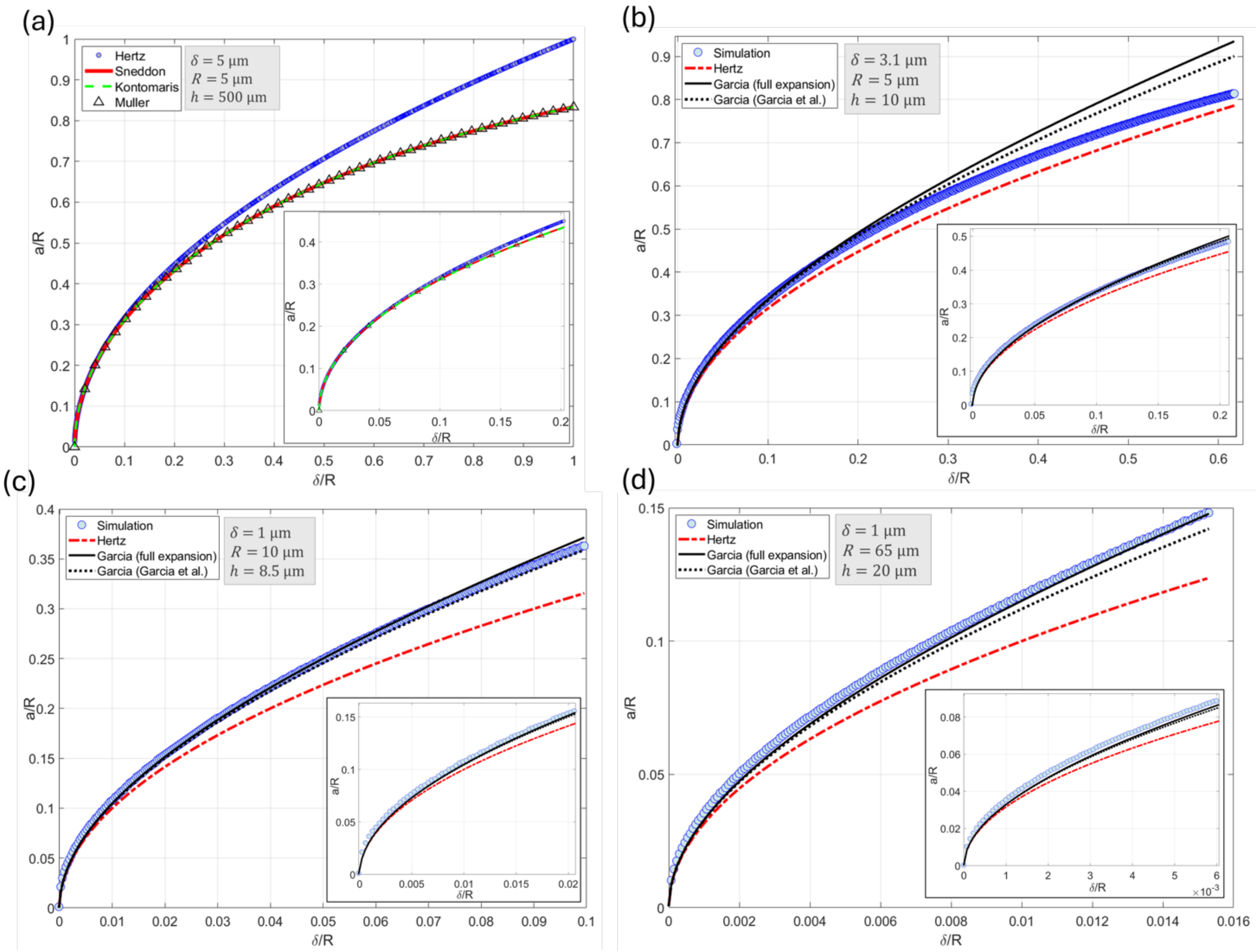

Figure S1. Evolution of the reduced contact radius *a/R* with the *δ/R* ratio for different simulated systems (bulk and thin films, spherical indenter) based on different contact mechanics models (Hertz and Sneddon models; Hertz + $LIC_{M/K}$ models; Hertz + $BEC_G$ models). The insets show an expanded view of the small $\delta/R$ values region. (a) Bulk systems. The contact radius including Muller's large indentation correction was calculated using Eq. S3 (see Note S1), whereas the radius including Kontomaris' correction is taken from Eq. 8 of Ref.[1]. The Sneddon radius is obtained from Eq. 4 in the main text. (b-d) Thin-Film systems, spherical indenter, different combinations of parameters *δ*, *R* and *h* leading to the same $\chi = 0.4$ (see Table 1 in the main text). The contact radius including Garcia's bottom effect correction up to the 1st order in $\chi$ was taken from Ref. [4] (Eq. S5 above), while the radius including Garcia's bottom effect correction up to the 5$^{th}$ order was calculated using Eq. S6 (see Note S2).

In bulk systems (a), the radii obtained using both Muller's and Kontomaris' large indentation corrections provide excellent agreement with the simulated data, as well as with the Sneddon model for the spherical indenter. In the thin film condition (b-d), the radii obtained using Garcia's bottom effect correction perform better than the Hertzian radius, as long as the large indentation effect is weak (i.e. for small $\delta/R$ values, compare (c,d) to (b)).

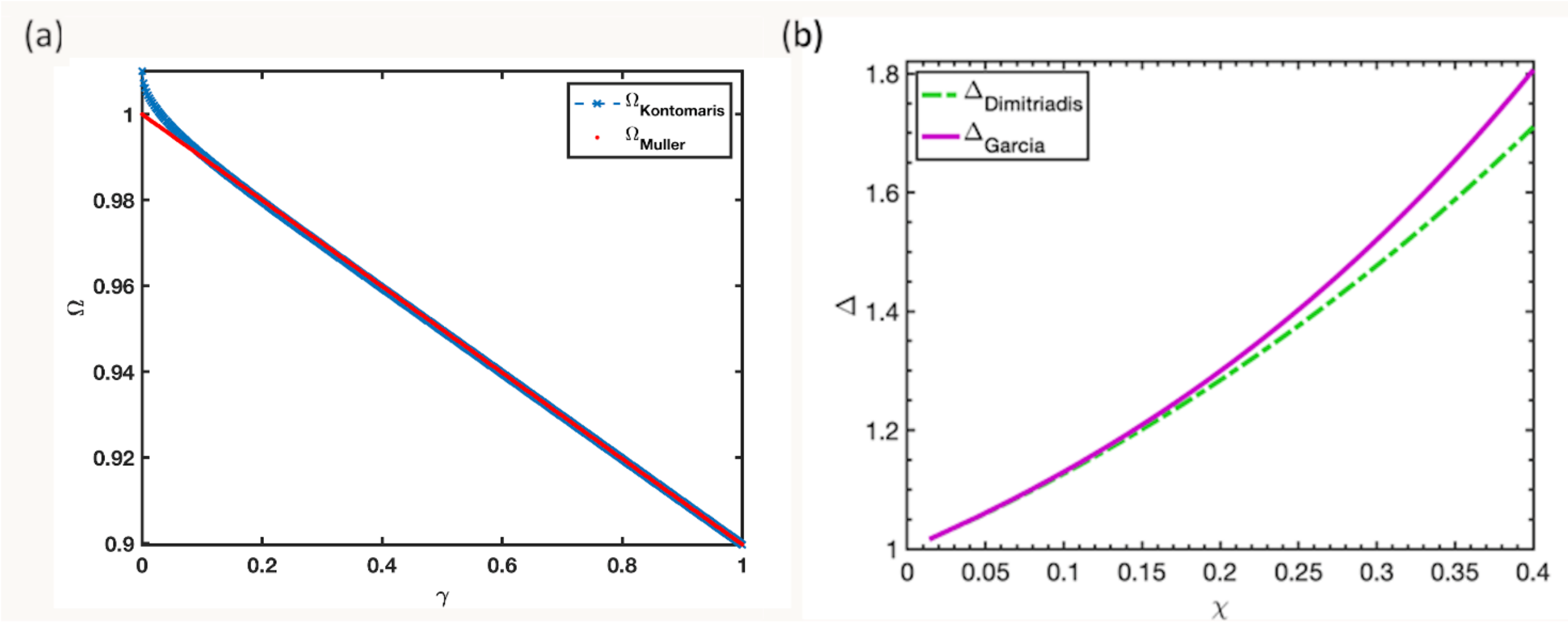

Figure S2. The correction functions $\Omega(\gamma)$ and $\Delta(\chi)$ for the Hertz model ($\gamma = \delta/R$, $\chi = a_{Hertz}/h = \sqrt{R\delta}/h$, $\nu = 0.5$) for a bonded sample in the explored ranges of the parameters: (**a**) LIC; (**b**) BEC.

## Note S3. Linearity of Tabor and Kalidindi effective stress on strain

Tabor defined effective stress $\sigma^*$ and strain $\varepsilon^*$ as[6]

$$\varepsilon^* = 0.2\frac{a}{R} = 0.2\frac{\delta}{a} \tag{S9}$$

$$\sigma^* = \frac{F}{\pi a^2} \tag{S10}$$

, where $a = \sqrt{R\delta}$ is the Hertzian contact radius and $F$ is the applied force, $\upsilon$ is the Poisson's ratio and $E$ is the Young's modulus. The prefactor 0.2 was empirically determined by Tabor [6]. At odd with the case of equivalent von Mises stress and strain, which are local quantities, Tabor stress and strain form a single pair of values representing the elastic response of the system as a whole. For a linear elastic material obeying Hertzian mechanics, it follows:

$$\sigma^* = \frac{20}{3\pi}\frac{E}{(1-\upsilon^2)}\varepsilon^* \tag{S11}$$

Kalidindi *et al.*[7] proposed an alternative definition of effective stress and strain:

$$\varepsilon^* = \frac{4}{3\pi}\frac{\delta}{a} \tag{S12}$$

$$\sigma^* = \frac{F}{\pi a^2} \tag{S13}$$

, from which it follows:

$$\sigma^* = \frac{E}{1-\upsilon^2}\varepsilon^* \tag{S14}$$

The effective Kalidindi's strain is approximately twice the one predicted by Tabor's formula; apart from a numerical factor, the scaling of the two stress-strain pairs is the same.

## Note S4. Linearity of equivalent von Mises stress on strain

In each simulation, a strain and stress field developes inside the deformed material. From the stress field, a scalar representing a pointwise stress state, called the equivalent (von Mises) stress field $\sigma_{\text{equiv}}$, may be computed. Similarly, the equivalent strain $\varepsilon_{\text{equiv}}$ exist. These two quantities are defined as below:

$$\sigma_{\text{equiv}} = \sqrt{\frac{(\sigma_1-\sigma_2)^2+(\sigma_2-\sigma_3)^2+(\sigma_3-\sigma_1)^2}{2}}$$

$$\varepsilon_{\text{equiv}} = \frac{1}{1+\nu}\sqrt{\frac{(\varepsilon_1-\varepsilon_2)^2+(\varepsilon_2-\varepsilon_3)^2+(\varepsilon_3-\varepsilon_1)^2}{2}} \qquad \text{(S15)}$$

, where $\nu$ is the Poisson's ratio, $\sigma_{\text{j}}$ and $\varepsilon_{\text{j}}$ are the principal stress and strain components, respectively, i.e. the eigenvalues of the stress and strain tensor. Since $\sigma_{\text{equiv}}$ and $\varepsilon_{\text{equiv}}$ are expressed in terms of the principal components, they are invariant under any change of base. For each simulation, ANSYS provides the maximum and the minimum equivalent stress and strain over time.

Starting from the generalized Hooke's law, in the regime of linear elasticity, it holds that

$$\sigma_{\text{equiv}} = E\varepsilon_{\text{equiv}}$$

where $E$ is the Young's modulus. The demonstration follows.

The starting point is the generalized Hooke's law

$$\varepsilon_{ik} = \frac{1}{9K}\delta_{ik}\sum_l \sigma_{ll} + \frac{1}{2\mu}\left(\sigma_{ik} - \frac{1}{3}\sum_l \sigma_{ll}\right)$$

, where $\sum_l \sigma_{ll} = Tr(\sigma)$ is the trace of the stress tensor, $K$ and $\mu$ are the uniform compression and shear modulus, respectively. By choosing a suitable base in which both $\varepsilon$ and $\sigma$ are diagonal, we compute the p-th principal value of the strain tensor (p=1,2,3) as

$$\varepsilon_p = Tr(\sigma)\left(\frac{1}{9K} - \frac{1}{6\mu}\right) + \frac{\sigma_p}{2\mu}$$

Therefore, the difference between the p-th and the b-th strain principal values reads

$$\varepsilon_p - \varepsilon_b = \frac{1}{2\mu}(\sigma_p - \sigma_b) \qquad \text{(S16)}$$

By inserting Eq. S16 in the definition of $\varepsilon_{\text{equiv}}$ (Eq. S15), one obtains:

$$\varepsilon_{\text{equiv}} = \frac{1}{2\mu(1+\nu)}\sigma_{\text{equiv}}$$

By noting that, by definition:

$$\nu = \frac{1}{2}\frac{3K - 2\mu}{3K + \mu}$$

, and

$$E = \frac{3K + \mu}{9K\mu}$$

, one finds:

$$\frac{1}{2\mu(1 + \nu)} = \frac{1}{E}$$

It follows:

$$\sigma_{\mathrm{equiv}} = E\varepsilon_{\mathrm{equiv}}$$

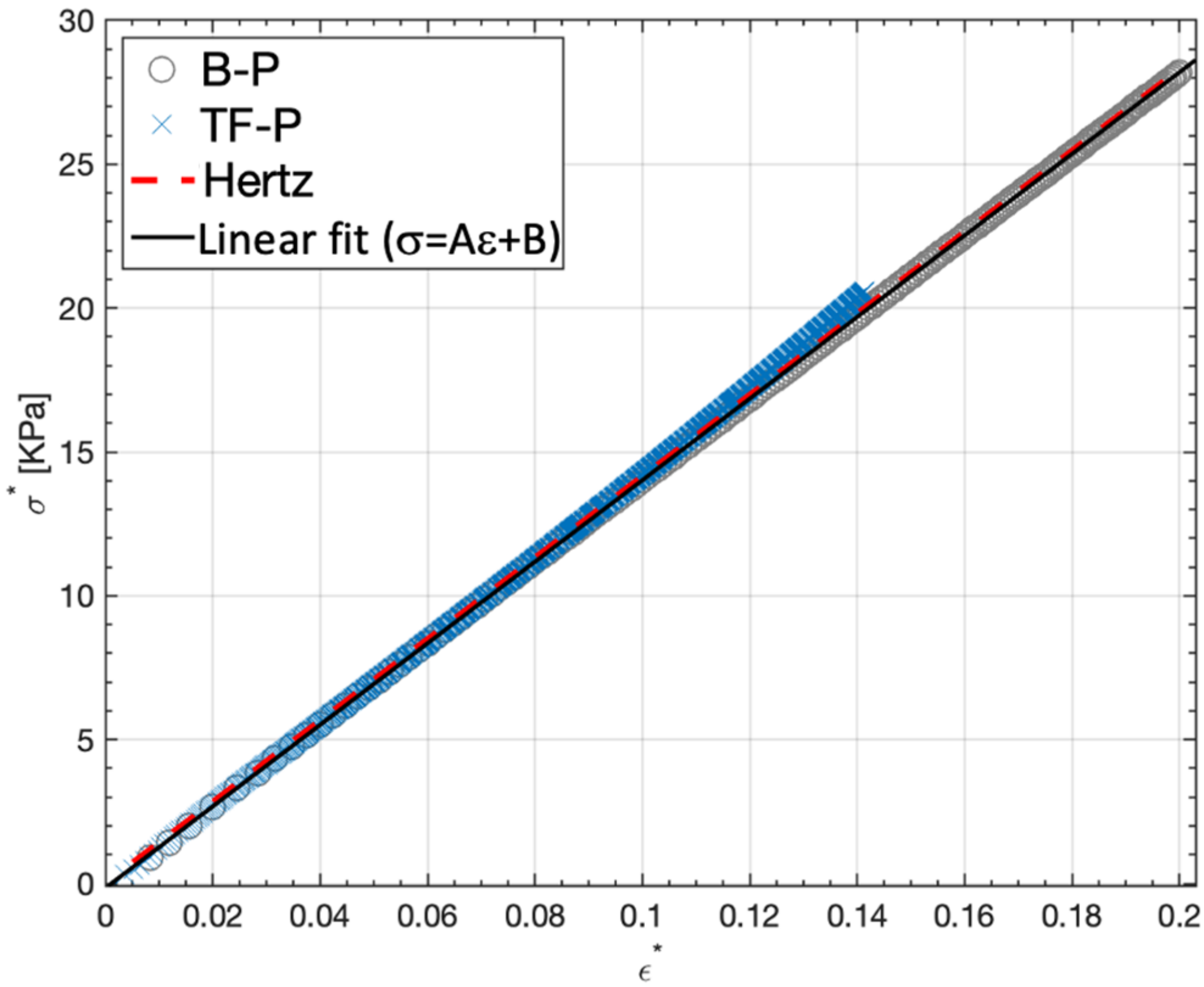

Figure S3. Tabor effective stress vs effective strain for the Bulk-Paraboloid (B-P) and Thin Film-Paraboloid (TF-P) configurations. Effective stress and strain have been calculated according to Eqs. S9, S10 (see Note S3). (o) Bulk-Paraboloid (B-P) simulated data. (**x**) Thin Film-Paraboloid (TF-P) simulated data, corrected using Garcia's formula $BEC_G$. (**---**) Effective stress-strain curve computed using the Hertzian contact radius. (—) Linear fit to the B-P simulated data.

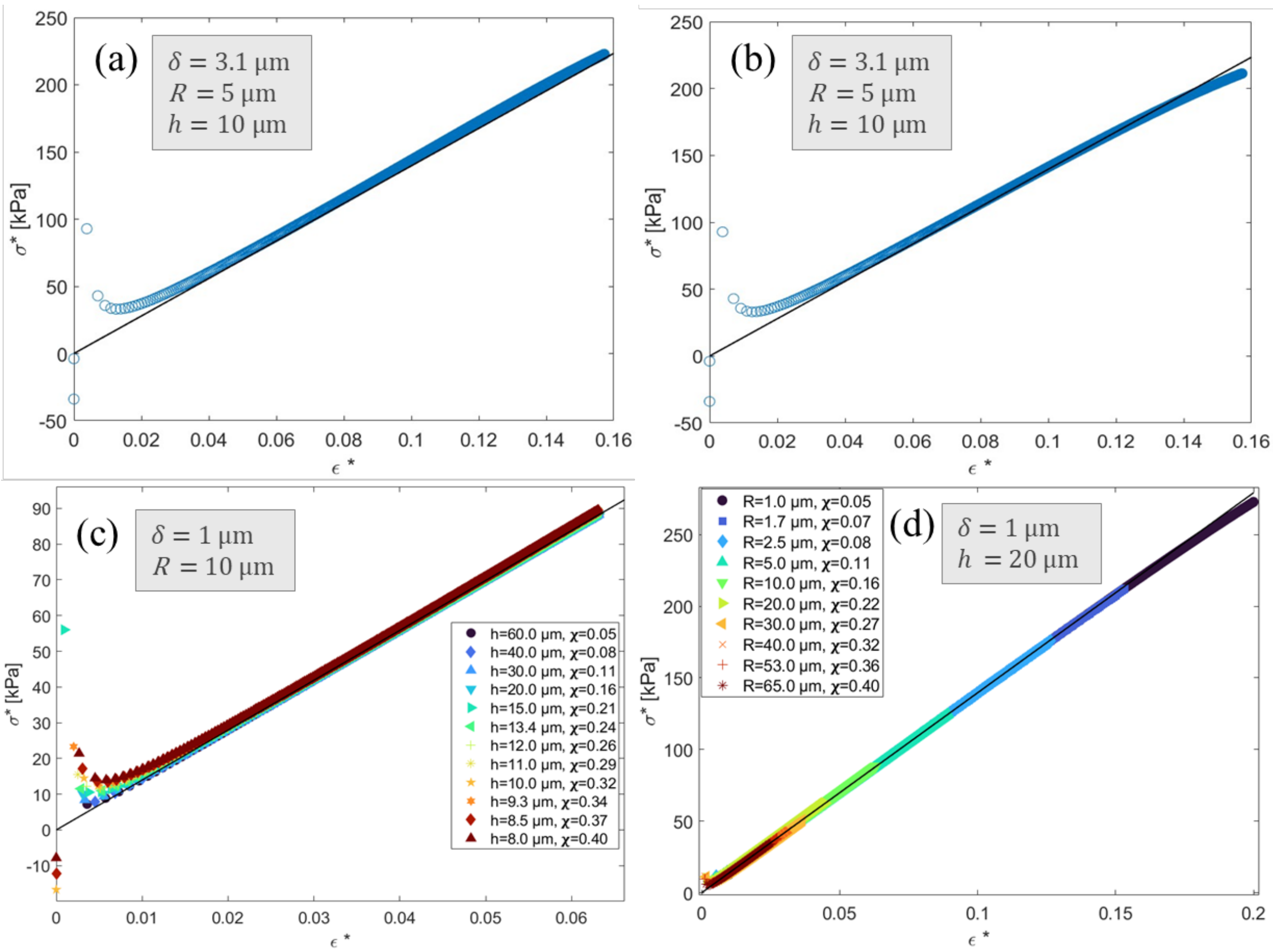

Figure S4. Effective stress vs effective strain plots of the Thin Film-Sphere (TF-S) systems (see Note S3). The effective stress and strains are computed using the Tabor[6] formulae (Eqs. S9, S10)[7] after applying BEC+LIC. At odd with the case of von Mises stress and strain (Figure S5f), which are local equivalent quantities, the effective Tabor stress and strain represent averages across the whole deformation volume, whose definition implies the specific Hertz model; the linearity of the effective stress vs effective strain relationship, therefore, must be tested after application of BEC and LIC. The force curves were corrected using Garcia's BEC and Muller's LIC, except in (a), where Dimitriadis' BEC was used, since it performs slightly better in the specific simulated conditions. (a) TF-S system corresponding to $\chi = 0.4$ corrected with Muller's LIC and Dimitriadis' BEC. (b) TF-S system corresponding to $\chi = 0.4$ corrected with Muller's LIC and Garcia's BEC. (c) TF-S systems in which the elastic medium thickness was varied. (d) TF-S systems where the tip radius was varied (see Table 1 in the main text). The solid black line represents Eq. S11, using the nominal Young's modulus of the elastic medium. The well-defined linear relations of the effective stress/strain curves confirm that in our simulations the linear elastic regime holds. The nonlinearity of the small-strain region can be attributed to the finite resolution (approx. 20-80 nm) of the simulation mesh, which affects both effective stress and strain values for $\varepsilon^* \lesssim 0.015$.

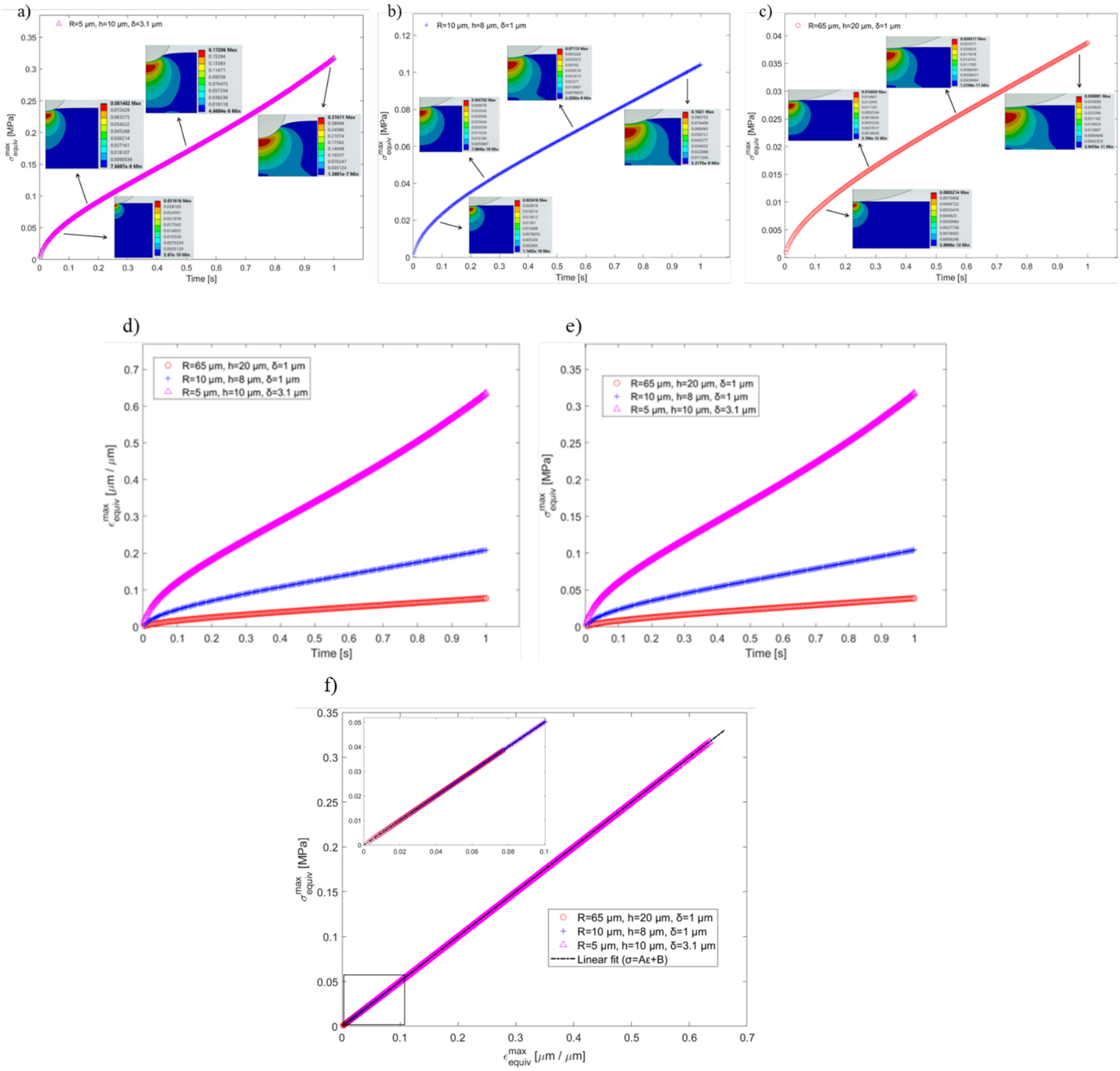

Figure S5. Maximum equivalent von Mises stress and strain for the Thin Film-Sphere (TF-S) systems. (**a-c**) Maximum equivalent von Mises' stress (see Note S4) versus fictitious time of the simulation for the three TF-S simulated systems (see Table 1 in the main text) with insets showing the geometry of the stress field developing in the elastic medium as the tip indents. The units of the insets are MPa. (**d-e**) maximum equivalent (von Mises) stress (same as in a-c) and strain of the three TS-S systems. (**f**) equivalent stress vs equivalent strain for the three TF-S systems, along with a linear fit $\sigma = A\varepsilon + B$ (black dashed line). The linear fit reproduces with excellent agreement the simulated data, providing $A = 0.5$ MPa, which is the nominal Young's modulus of the simulated material, and $B \approx 0$, as expected. It is worth stressing that the fitted equivalent stress and strain are raw data taken from the simulations, hence neither BEC nor LIC corrections were applied. These results are consistent with the fact that the linear elastic behaviour holds also in confined geometries, irrespective to the boundary conditions. The fact that the curves follow a linear trend further confirms that the linear elastic regime holds in all our simulations.

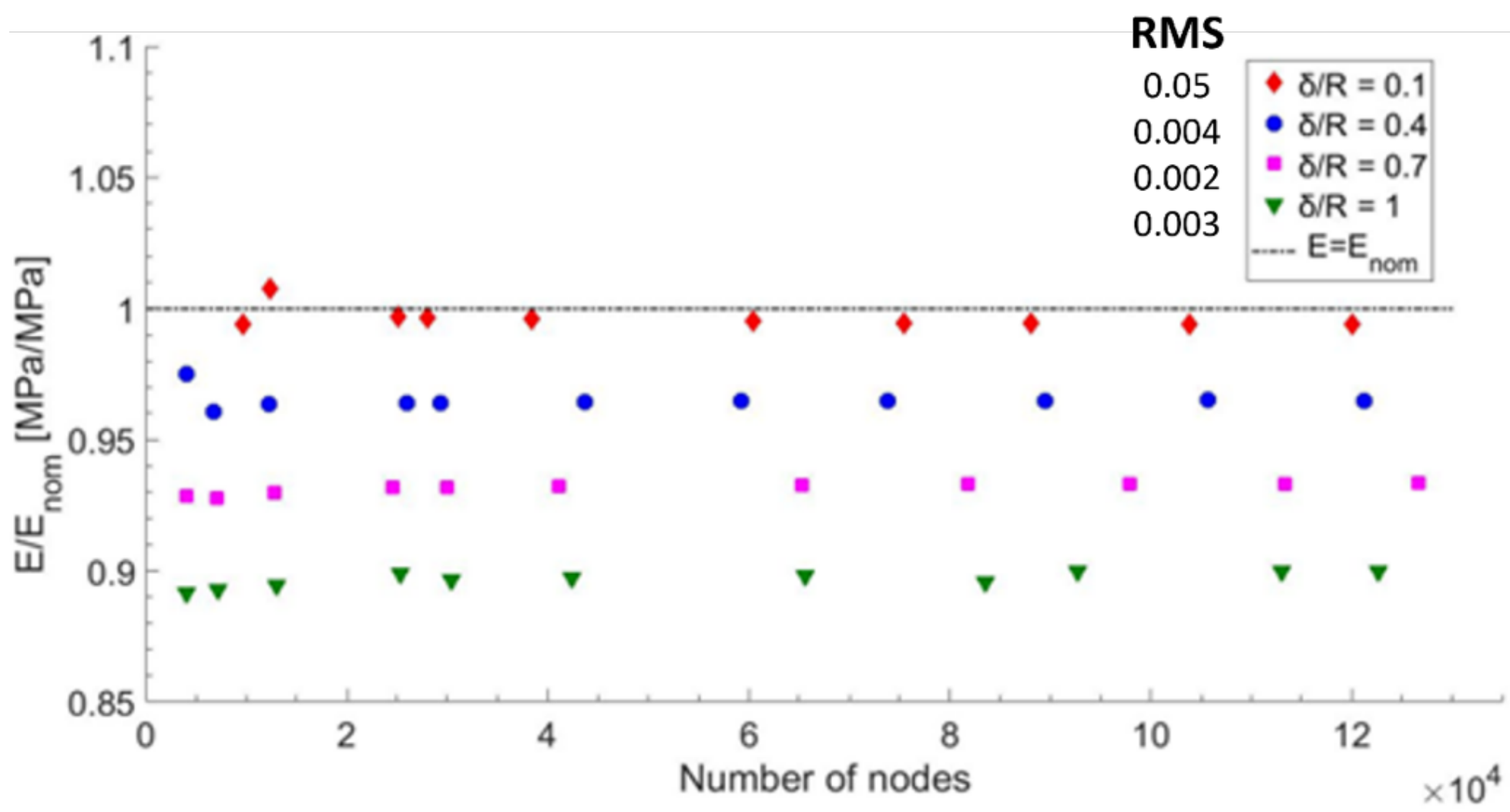

Figure S6. Impact of the number of nodes in the mesh of finite element simulations. The Young's modulus $E$ was measured fitting the Hertz model to force curves simulated with the Bulk-Sphere (B-S) model, while $E_{nom} = 0.5$ MPa is the nominal value used in the simulations. As the $\delta$/R ratio increases and the contact departs from the hertzian model (large indentation effect), the Young's modulus decreases, irrespective of the number of nodes. The dashed line refers to the ideal hertzian contact. The root mean square (RMS) values of each series are reported.

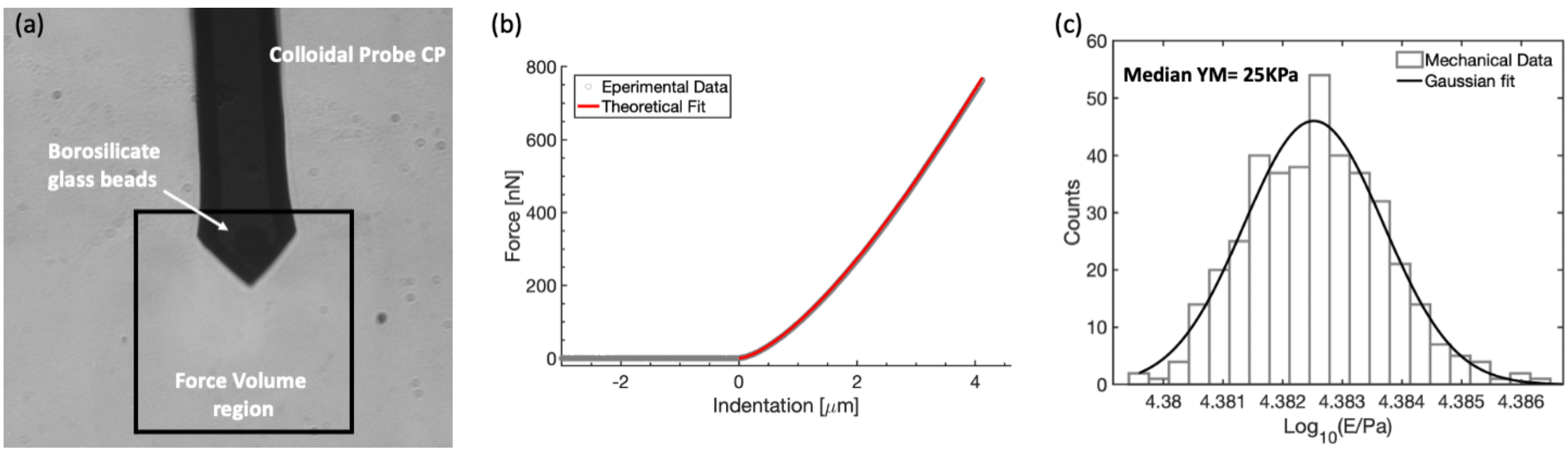

Figure S7. Schematics of the applied nanomechanical protocol. (a) Selection of the region of interest to acquire a Force Volume on the sample via optical microscopy. (b) Force curves extracted from the FV (already aligned and rescaled) with the Hertzian fit highlighted. (c) Histogram of the YM values measured from each FC in semilog10 scale. The gaussian fit of the YM distribution in log scale provides the median YM value as: $E_{median}$=10log$_{10}$(E/Pa)|best, where log$_{10}$()|best is the centre of the gaussian curve.

## Bibliography

[1] S. V Kontomaris, A. Malamou, *Eur J Phys* **2021**, *42*, 025010.

[2] P. Müller, S. Abuhattum, S. Möllmert, E. Ulbricht, A. V. Taubenberger, J. Guck, *BMC Bioinformatics* **2019**, *20*, DOI 10.1186/s12859-019-3010-3.

[3] W. C. Oliver, F. R. Brotzen, *J Mater Res* **1992**, *7*, 613.

[4] P. D. Garcia, R. Garcia, *Biophys J* **2018**, *114*, 2923.

[5] E. K. Dimitriadis, F. Horkay, J. Maresca, B. Kachar, R. S. Chadwick, *Biophys J* **2002**, *82*, 2798.

[6] D. C. Lin, D. I. Shreiber, E. K. Dimitriadis, F. Horkay, *Biomech Model Mechanobiol* **2009**, *8*, 345.

[7] S. R. Kalidindi, S. Pathak, *Acta Mater* **2008**, *56*, 3523.